\documentclass{article}

\usepackage{PRIMEarxiv}
\usepackage[utf8]{inputenc} 
\usepackage[T1]{fontenc}    
\PassOptionsToPackage{hyphens}{url}
\usepackage{hyperref}       
\usepackage{booktabs}       
\usepackage{amsfonts}       
\usepackage{nicefrac}       
\usepackage{microtype}      
\usepackage{lipsum}
\usepackage{fancyhdr}       
\usepackage{graphicx}       
\usepackage{subcaption}
\graphicspath{{media/}}     
\usepackage{multirow}
\usepackage{makecell}

\usepackage{tabularx}
\usepackage{booktabs}
\usepackage{ragged2e}

\usepackage{enumitem}

\usepackage{verbatim}       
\usepackage{amsmath}

\usepackage{float}

\usepackage{multirow}

\usepackage{soul}

\usepackage{longtable}      
\usepackage{listings}
\usepackage{colortbl}       

\usepackage{tcolorbox}
\tcbuselibrary{listings,skins,breakable}

\lstdefinestyle{searchstyle}{
    language=searchquery,
    basicstyle=\ttfamily\small,
    keywordstyle=\color{blue}\bfseries,
    stringstyle=\color{teal!60!black},
    breaklines=true,
    columns=fullflexible
}

\newcommand{\sunny}[1]{\textcolor{blue}{\textbf{Sunny:} #1}}

\newcommand{\limingD}[1]{\textcolor{brown}{\textbf{LimingD:} #1}}

\title{A Structured Approach to Safety Case Construction for AI Systems}

\author{
  Sung Une Lee, Liming Zhu, Md Shamsujjoha, Liming Dong, Qinghua Lu, Jieshan Chen, Lionel Briand*\\
  Data61, CSIRO, Australia\\
  \texttt{\{firstname.lastname\}@data61.csiro.au}
  \\
   *University of Ottawa, Canada\\
  *\texttt{lbriand@uottawa.ca}
  }

\begin{document}
\maketitle

\begin{abstract}
Safety cases, structured arguments that a system is acceptably safe, are becoming central to the governance of AI systems. Yet, traditional safety-case practices from aviation or nuclear engineering rely on well-specified system boundaries, stable architectures, and known failure modes. Modern AI systems, such as generative and agentic AI, are the opposite. Their capabilities emerge unpredictably from low-level training objectives, their behaviour varies with prompts, and their risk profiles shift through fine-tuning, scaffolding, or deployment context. This study examines how safety cases are currently constructed for AI systems and why classical approaches fail to capture these dynamics. This study introduces comprehensive taxonomies for AI-specific claim types (assertion-based, constraint-based, capability-based), argument types (demonstrative, comparative, causal/explanatory, risk-based, and normative), and evidence families (empirical, mechanistic, comparative, expert-driven, formal methods, operational/field data, and model-based). It then proposes a reusable safety-case template, each of which follows a predefined structure of claims, arguments, and evidence tailored for AI systems. Each template is illustrated by end-to-end patterns that address distinctive challenges, such as evaluation without ground truth, dynamic model updates, and threshold-based risk decisions. The result is a systematic, composable, and reusable approach to constructing and maintaining safety cases that are credible, auditable, and adaptive to the evolving behaviour of generative and frontier AI systems.
\end{abstract}

\keywords{AI \and AI Safety Case \and AI Safety Case Taxonomy \and AI Safety Case Template \and AI Safety Case Pattern}

\section{Introduction} \label{sec:introduction}


Safety cases have become a central assurance artefact across safety-critical engineering, structured arguments supported by evidence, intended to demonstrate that a system is acceptably safe for a specific purpose and operating context~\cite{kelly1999arguing, bloomfield2009safety}. 
However, artificial intelligence (AI) systems, including frontier, generative, and agentic AI systems, fundamentally differ from the deterministic, specification-driven systems for which safety cases were originally designed. Their capabilities are not engineered line by line but discovered after training. Their risks evolve through interactions, fine-tuning, and context. In addition, their evaluation often proceeds without a fixed ground truth. Consequently, the logic of assurance for AI should be discovery-driven, continually updated, and capable of integrating empirical, statistical, and abductive reasoning alongside classical deductive forms.


The value of a safety case lies not only in regulatory compliance or as a governance tool, but also in being an epistemic discipline that stabilises ethical reasoning under uncertainty. Discussions of AI safety and ethics often operate largely in a pre-harm, anticipatory mode~\cite{Konigs2025Negativity}, where severity is inferred rather than observed. While this is necessary, such analysis is often conducted with limited empirical grounding, making it difficult to specify risk severity, compare it with existing systems, or define acceptable risk thresholds.
The safety-case approach is well-suited to addressing these challenges. By grounding ethical and safety concerns in explicit claims, evidence, and well-defined comparators or baselines, AI safety cases support consistent risk assessment, review, and justification across different systems and contexts.

While early work has begun to adapt safety-case thinking to frontier AI~\cite{buhl2024safety, carlan2024dynamic, clymer2025example}, the field still lacks a coherent, reusable structure that practitioners can apply across model types, lifecycle stages, and regulatory regimes. Existing examples tend to focus on isolated instances such as dynamic updating, misuse safeguards, or autonomous-system templates, without an overarching synthesis that ties together claim formulation, argument structure, and evidence design. Moreover, current practices rarely capture the iterative process by which AI developers discover new model capabilities and failure modes through evaluation, stress testing, and adversarial probing~\cite{bommasani2021opportunities}. 
These discoveries continually reshape the scope of assurance itself.

This study develops a comprehensive taxonomy of AI safety cases and reusable safety-case templates for AI systems, addressing the limitations of existing assurance approaches. 
The key contributions are fourfold.
\begin{itemize}
    \item \textbf{Characterisation of modern AI safety cases.} The study first analyses how safety cases are currently constructed for AI systems, contrasting them with classical engineering approaches and identifying distinctive features, including capability discovery, the absence of ground truth, continuous evolution, and threshold-based decision-making.
    \item \textbf{Taxonomies for reusable safety-case design.} It introduces AI-specific taxonomies for \textit{Claim types}, including assertion-based, constrained-based, capability-based claims, \textit{Argument types} including demonstrative, comparative, causal/explanatory, risk-based, and normative arguments, and \textit{Evidence families} such as empirical, mechanistic, comparative, expert-driven, formal methods, model-based, and operational/field data. Each is linked through explicit reasoning logics (deductive, inductive, abductive, statistical, and analogical).\\
    The taxonomies introduced in this study are descriptive and compositional, not mutually exclusive classification schemes. Claim types, argument types, and evidence families are intended as orthogonal lenses on safety reasoning, rather than disjoint buckets into which safety cases must be partitioned. In real AI safety cases, overlap across categories is expected and legitimate.
    \item \textbf{Template and pattern library.} It proposes a library of reusable templates for constructing AI safety cases, illustrated through end-to-end patterns that address unique AI challenges such as safety justification through \textit{discovery-driven} evaluation, \textit{marginal-risk reasoning} without ground truth, \textit{continuous evolution} for dynamic update and redeployment management, and \textit{threshold-based} risk acceptance.
    \item \textbf{Integration with dynamic assurance.} Building on emerging ideas such as Dynamic Safety-Case Management Systems (DSCMS) and Checkable Safety Arguments~\cite{carlan2024dynamic}, the study embeds these templates within a continuous evaluation pipeline, linking safety claims to live metrics and governance artefacts.
\end{itemize}

Existing AI safety case approaches typically provide domain-specific instantiations (e.g., cyber inability templates), governance-oriented guidance for frontier AI, or concepts for dynamic safety management. However, they focus on specific assurance elements and do not jointly provide an integrated CAE taxonomy spanning claims, arguments, and evidence, a reusable template meta-model, and a composable pattern library that can be instantiated across recurring AI assurance scenarios.
The contributions of this study establish a composable, auditable, and reusable foundation for constructing safety cases that remain credible under the uncertainty, discovery, and rapid change inherent to modern and emerging AI systems.
A detailed comparison with representative approaches is provided in Table \ref{tab:positioning} (Section \ref{sec:background}).

To illustrate the practical implications of this study, we examine a real-world case study involving an AI-based tender evaluation system in government. This case study demonstrates the practical application of the proposed approach, highlighting its potential to enhance governance and assurance processes. It also provides concrete insights that inform decision-making and guide the development of effective, responsible AI solutions.



The remainder of the paper proceeds as follows.

Section \ref{sec:background} provides background on AI safety cases, including an in-depth analysis of how they diverge from traditional engineering examples, highlighting the distinctive challenges of discovery-driven evaluation and evolving system boundaries. 
We then present the methodology of this study in Section \ref{sec:methodology}.
Section \ref{sec:taxonomy} introduces the reusable safety-case templates and details the proposed taxonomies of claim, argument, and evidence (CAE) types. Section \ref{sec:pattern} presents end-to-end patterns that illustrate how these templates address characteristic AI assurance problems, such as marginal risk in the absence of ground truth and continuous model evolution. Section \ref{sec:case study} provides a case study to show how the study can be applied in the real-world context.
In Section \ref{sec:limitation}, we discuss current limitations of this study and future work. 
We then conclude this study in Section \ref{sec:concusion}.

\section{Background} \label{sec:background} 

In this section, we introduce core AI safety case concepts, including the Claims–Arguments–Evidence (CAE) approach, taxonomy, template, and pattern, present an overview of the AI safety ecosystem pipeline involving different stakeholders, and examine the transition from traditional engineering safety cases to AI-specific approaches.

\subsection{AI safety case concepts}


\textbf{Safety Case.} 
A safety case is a documented, structured argument asserting that a system is acceptably safe for a given context of operation, supported by evidence~\cite{mod2007safety,habli2025big}. For AI systems, especially high-capability or frontier AI, the safety case aims to show that the system's deployment or operation does not pose unacceptable risk.

\begin{figure}[t]
  \centering
  \includegraphics[width=.8\textwidth]{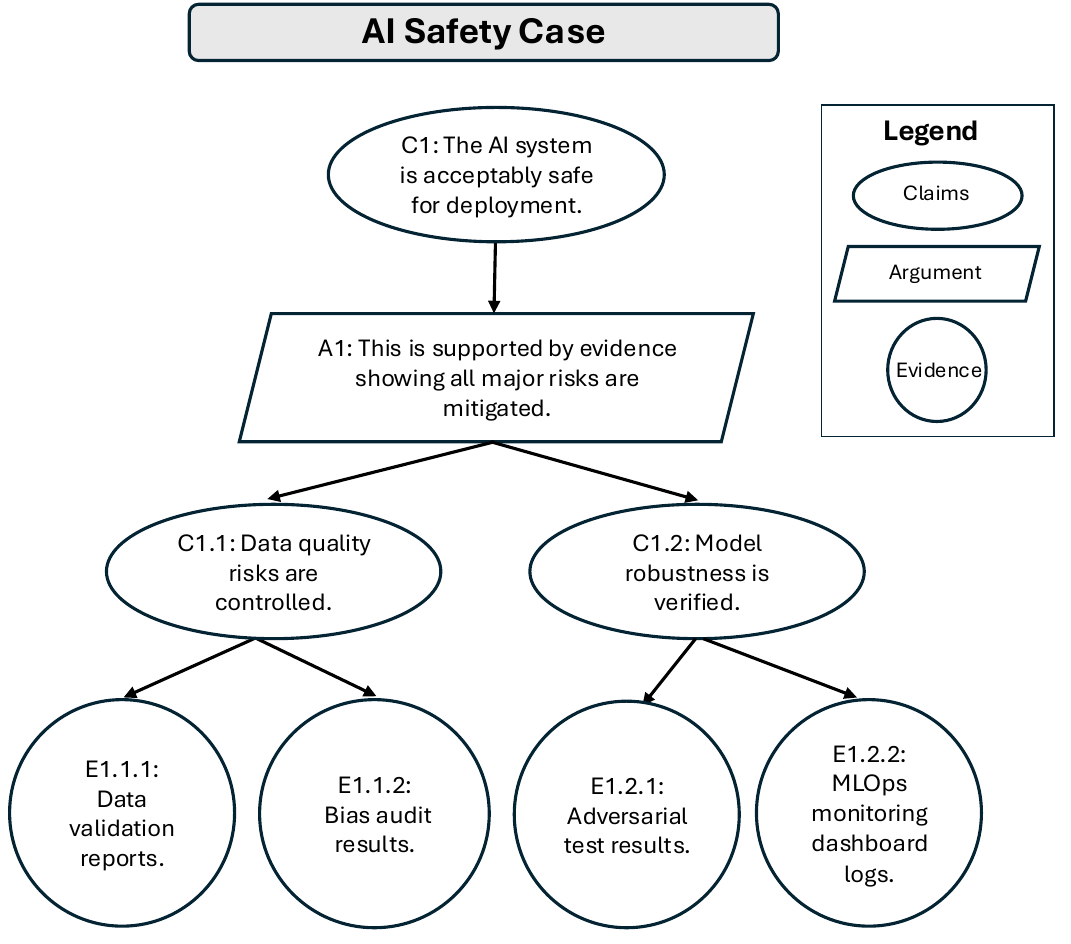}
  \caption[]{Claims, Arguments, Evidence (CAE) example}
  \label{fig:CAE concepts}
\end{figure}

\textbf{\\ \\Claims–Arguments–Evidence (CAE).}
Recent works have developed safety cases using the CAE structure~\cite{bloomfield2021templates,goemans2024template}. For example, Goemans et al.~\cite{goemans2024template} propose a safety-case template for cyber inability that explicitly uses CAE to make safety arguments coherent and explicit. The CAE structure (see Figure~\ref{fig:CAE concepts})~\cite{caeframework2024} emphasizes:
\begin{itemize}
    \item \textbf{Claim:} A claim is a true/false statement about a property of a particular object. A claim is just what you might consider to be from common usage of the term; an idea that someone is trying to convince someone else is true. An example claim could be what you are asserting (e.g., "the model cannot perform X harmful capability").
    \item \textbf{Argument:} An argument is a rule that provides the bridge between what we know or are assuming (sub-claims, evidence) and the claim we are investigating. In other words, it refers to the justification linking a specific claim to a specific piece of evidence~\cite{adelard2024ascad, korbak2025sketch}. The reasoning or structure showing why the claim holds (e.g., decomposition into sub-claims, use of proxies, evaluations).
    \item \textbf{Evidence:} Evidence is an artefact that establishes facts that can be trusted and lead directly to a claim. E.g., data, test results, evaluations, or simulation results that support the argument.
\end{itemize}

\begin{figure}[t]
  \centering
  \includegraphics[width=1\textwidth]{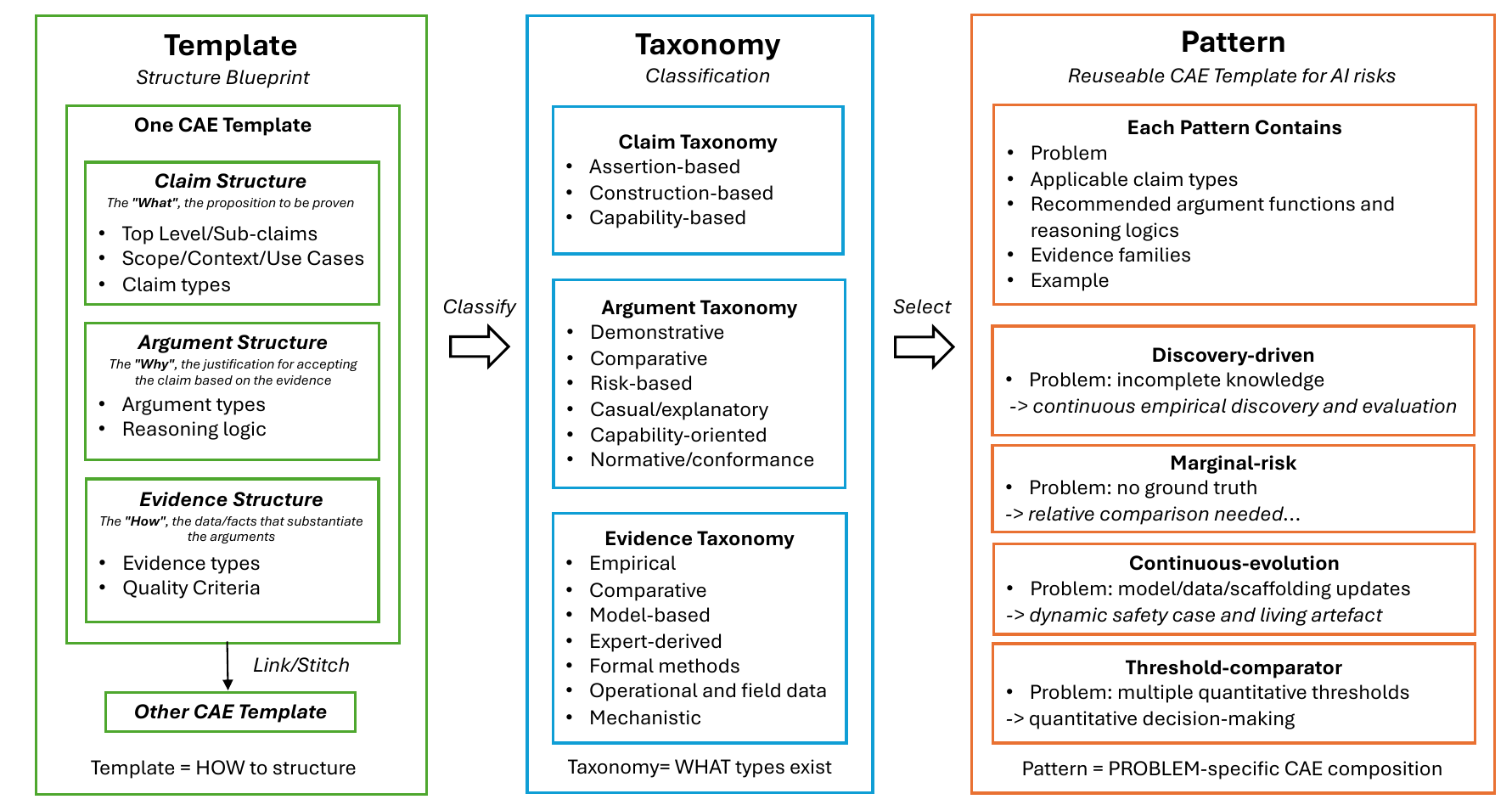}
  \caption[]{AI safety case template, CAE taxonomy, and pattern}
  \label{fig:TTP}
\end{figure}

In this study, the CAE taxonomy provides a unified classification of claim, argument, and evidence types, yet classification alone does not show how these elements should be assembled into a working safety case. Templates address this gap by offering structured blueprints that operationalise the taxonomy, and patterns further extend these templates by composing them into reusable solutions for recurring AI-specific risks.
Figure~\ref{fig:TTP} illustrates the relationships among taxonomy, template, and pattern in AI safety case construction.

\textbf{Safety Case Taxonomy.} A \emph{taxonomy} denotes a categorical classification scheme for safety case elements, organizing the different types of claims, argument structures, and evidence categories that may appear in an AI safety case. Research on safety-case taxonomies spans foundational argument structures, frequent evidence types, and emerging frontier-AI–specific argument types. Kelly~\cite{kelly1999arguing} classifies reusable argument structures into top-down, bottom-up, and general construction patterns, introducing formal notions of claim decomposition and argument strategies. Nair et al.~\cite{nair2014evidence} provide a taxonomy of evidence types for safety certification, including Hazards Cause Specification and Testing Results. Buhl et al.~\cite{buhl2024safety} identify four core safety-case components and introduce frontier-AI specific argument types, such as directly risk-based arguments and inability arguments, reflecting the shift from traditional failure-based reasoning to arguments about model behaviour and emergent capabilities.

However, existing taxonomies ~\cite{clymer2024justifysafety, wasil2024affirmativesafety, balesni2024evaluations} face several limitations that hinder their application to contemporary AI systems.
First, the absence of standardized taxonomies leads to inconsistent terminology and incomparable safety claims across organizations; what one team calls a "robustness claim" may differ substantially from another's interpretation. Second, existing taxonomies focus on isolated aspects (patterns, argument, evidence types, or high-level components) rather than providing an integrated view of the complete CAE structure. Third, prior work has not adequately addressed AI-specific challenges, including the emergence of capabilities, continuous evolution, and context-dependent behaviours that distinguish AI systems from traditional safety-critical systems.
We introduce our taxonomy to address critical gaps in current AI safety practices.
Building on a review of existing work, our CAE taxonomy provides a systematic basis for analyzing the primary AI safety case papers (\hyperref[AppendixB]{Appendix B}) we studied, enabling us to identify which claim types, argument structures, and evidence categories are most prevalent in current practice.

    
    

\textbf{Safety Case Templates.} 
A \emph{template} denotes a reusable structured blueprint for constructing a concrete safety case in a particular context or domain. While the taxonomy classifies element types, a template specifies how selected instances of these elements are arranged and how multiple CAE chains are linked or stitched to form a coherent assurance argument. 
%

Previous work on safety case templates originates from safety-critical domains such as autonomous vehicles. One influential work is the comprehensive set of templates proposed by Bloomfield et al.~\cite{bloomfield2021templates}, which outlines CAE-based templates for autonomous systems, including requirements templates, hazard-analysis templates, safety-monitor templates, and ML-sensor templates. 
Recently, template-based approaches have extended into frontier AI. Goemans et al.~\cite{goemans2024template} have developed a cyber-inability safety case template that integrates model capability evaluation as a first-class component, a methodological shift required by the opacity of modern AI systems, where behavioural evidence must be empirically derived rather than analytically proven. 

However, existing template work ~\cite{ bloomfield2021templates,goemans2024template, thomas2021template} has limited coverage of diverse claim, argument, and evidence categories. 
We present templates for each CAE element: \emph{Claim} is structured from top-level claims to sub-claims, each of which specifies the key objectives that must be addressed to demonstrate an AI system's safety in a specific operational context. \emph{Argument} provides diverse breakdown types that capture the justification and reasoning for accepting the claim given the evidence. \emph{Evidence} presenting the data and facts that substantiate the arguments, with diverse evidence types and quality criteria to ensure robustness, traceability, and sufficiency of the safety case. Our templates provide a structured lens for analyzing existing AI safety cases and guide practitioners in systematically constructing safety cases.

\textbf{Safety Case Patterns.} A \emph{pattern} denotes a reusable argument fragment or reasoning structure that captures a proven approach to addressing specific safety problems. Unlike templates, which provide complete structural blueprints, patterns are modular components that can be instantiated and combined across multiple contexts, embodying recurring reasoning strategies such as red-team/blue-team evaluation or defense-in-depth arguments.
Research on safety case patterns has evolved from traditional systems to AI applications. Kelly et al.~\cite{KellyM97Patterns} and Alexander et al.~\cite{Alexander2007Patterns} introduce early pattern-based methods for structuring safety arguments and supporting cross-project reuse. With the rise of ML, Wozniak et al.~\cite{WozniakCAP20PatternML} propose patterns covering data quality, model training, and runtime monitoring, while Kaur et al.~\cite{KaurICSL20PatternsDNN} extend these to deep learning with robustness testing and neural-network evaluation. Porter et al.~\cite{porter2024principles} propose principles-based patterns that frame ethical principles (fairness, transparency) as top-level claims, demonstrating broader assurance objectives beyond technical safety.

We have found that existing patterns ~\cite{KellyM97Patterns, Alexander2007Patterns, WozniakCAP20PatternML, KaurICSL20PatternsDNN} remain focused on specific traditional software or ML system challenges without coverage of frontier AI risks such as emerging capabilities, the absence of ground truth, continuous evolution behaviours, and a lack of explicit connections to the broader CAE framework. We introduce four structured AI safety case patterns (Section ~\ref{sec:pattern}) to address emerging AI risk challenges.
    
    
    


\subsection{Overview of AI safety cases ecosystem pipeline}



\begin{figure}[t]
  \centering
  \includegraphics[width=1\textwidth]{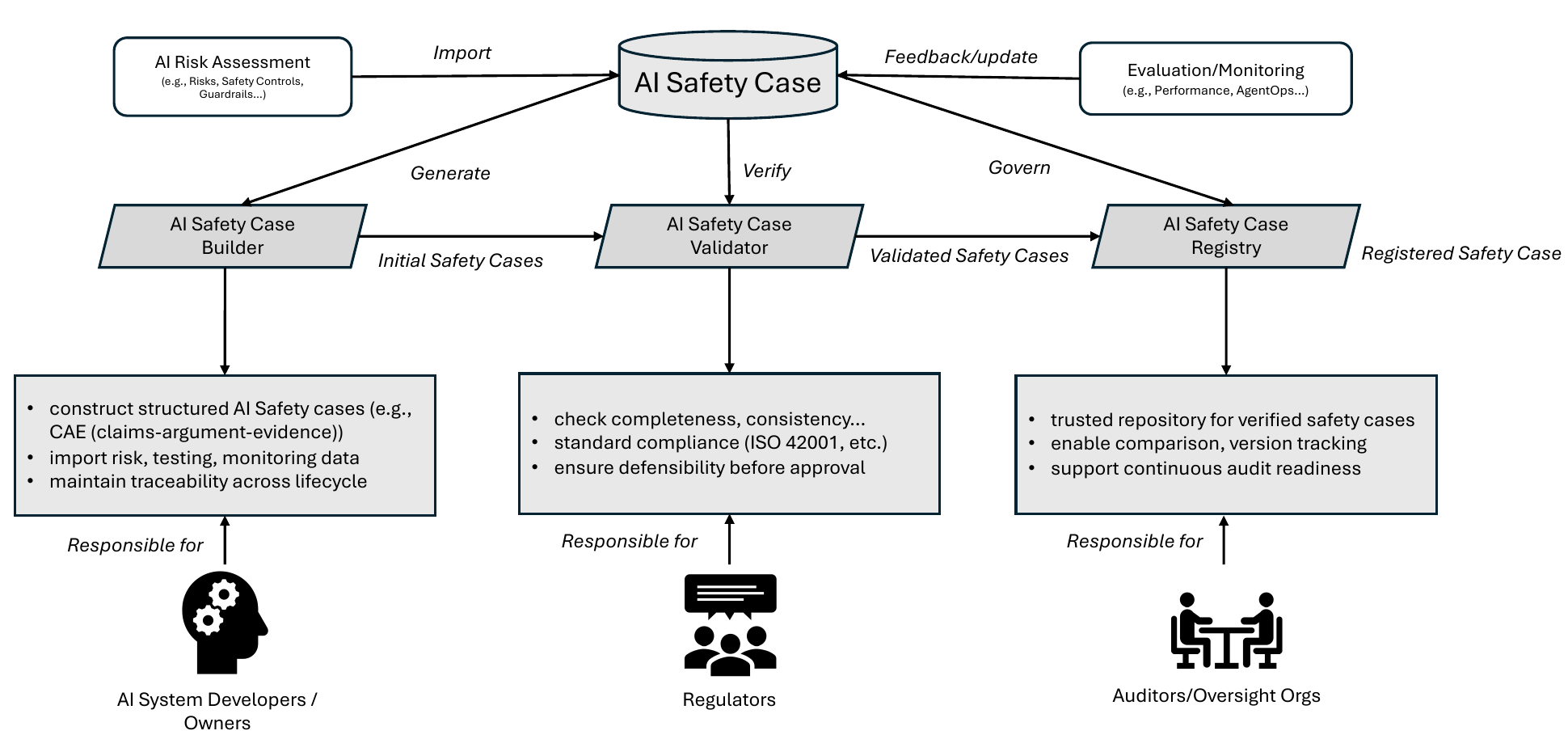}
  \caption[]{Overview of AI safety cases ecosystem pipeline}
  \label{fig:pipeline}
\end{figure}

To situate AI safety cases within real operational contexts, 
Figure~\ref{fig:pipeline} presents an overview of the pipeline that supports how 
AI safety cases are created, assured, and governed by different stakeholders in 
practice. In real deployments, safety cases are not static documents produced by 
developers alone; they form a shared artefact that connects multiple 
organisations with distinct responsibilities, including AI system developers and 
owners, government regulators, and independent oversight bodies. Each of these 
actors rely on the safety case for different purposes: developers to 
demonstrate risk-informed design and testing, and have regulators verify compliance 
and defensibility before approval, and audit organisations to maintain 
long-term accountability across versions.

Figure~\ref{fig:pipeline} therefore maps coordinated components' needs into a 
coherent pipeline centred on the evolving AI safety case, which is continuously 
updated with inputs from risk assessments, testing, evaluation, and operational 
monitoring~\cite{NOPSEMA2025safetycase,dong2024agentops2}. These three coordinated components, the \textit{AI Safety Case Builder}, 
\textit{AI Safety Case Validator}, and \textit{AI Safety Case Registry}, support 
this lifecycle by generating initial safety cases, verifying their completeness 
and compliance, and preserving validated versions for ongoing oversight. 
\begin{itemize}
    \item \textbf{AI Safety Case Builder/plugins:} help AI developers and system owners construct structured, versioned safety cases across the AI lifecycle. At design time, the builder imports outputs from risk assessments~\cite{buhl2024safety} to establish the structure of claims and required evidence, and later integrates results from testing, evaluation, and monitoring. Each design or testing iteration automatically updates the safety case, maintaining traceability between identified risks, implemented controls, and empirical results that demonstrate AI system safety and reliability.

    \item \textbf{AI Safety Case Validator:} used by government assurance bodies or accredited third parties to examine the integrity, completeness, and consistency of submitted safety cases. The Validator could check alignment with standards such as the Australian AI Safety Standard~\cite{VAISS_V1} or ISO 42001~\cite{ISO42001_2023}, verify the authenticity of evidence, and assess whether controls remain effective over time. It ensures that the safety case is auditable, transparent, and defensible before approval or deployment in critical contexts.
 
    \item \textbf{AI Safety Case Registry:} acts as a trusted repository for storing and sharing verified safety cases across agencies and sectors. It would enable oversight organisations~\cite{fairlyAI2025trust} to compare systems, trace changes, and support continuous audit readiness.
\end{itemize}
The \textit{Builder}, \textit{Validator}, and \textit{Registry} would form a pipeline for generating, verifying, and governing AI safety cases, making AI assurance a structured, evidence-based, and collaborative process among different stakeholders, e.g., developers, regulators, and auditors.

\subsection{From traditional engineering to AI safety cases}

Having introduced the core CAE concepts and the broader safety-case ecosystem, we now examine why traditional engineering approaches to safety assurance cannot be directly applied to AI systems. This contextualises the need for the AI-specific taxonomies, templates, and patterns developed in this study.

\textbf{The foundations of classical safety cases.}
In traditional engineering domains such as aerospace, rail, and nuclear power, safety cases serve as structured justifications that a system is acceptably safe for its intended use 
~\cite{kelly1999arguing, bloomfield2009safety}. The underlying logic is relatively stable: systems are designed to precise specifications, hazards can be systematically identified through techniques such as Fault Tree Analysis or Failure Mode and Effects Analysis, and risk is managed by design-time redundancy, physical containment, and procedural barriers. In these cases, the safety argument typically follows a deductive and hierarchical form: top-level claims (e.g., "The aircraft control system is acceptably safe") are decomposed into subclaims about components, processes, and operating conditions, supported by deterministic evidence such as design verification, test coverage results, and formal proofs.

These cases depend on predictability: the same inputs produce the same outputs, and the behaviour of the components is well understood. Assurance rests on design sufficiency (the system behaves as specified) and evidence completeness (all hazards have been analysed). Once the system is validated against fixed requirements, safety justification largely stabilises until the next design change or certification cycle.

\textbf{Why AI systems break this assurance model.}
AI systems, particularly generative and frontier AI models, disrupt nearly all these premises. They are not engineered line-by-line to a full behavioural specification, but are trained on massive, uncurated data using a low-level objective, such as next-token prediction. Their performance and risks are emergent properties of both data and training scale. Consequently, their capabilities are discovered after training rather than defined before it~\cite{bommasani2021opportunities}.

While traditional safety cases already address probabilistic risk models, socio-technical systems, human-in-the-loop control, partial observability, and operational drift, AI systems introduce distinctive characteristics that render traditional safety-case logic insufficient on their own:

\begin{itemize}
    \item \textbf{Post-deployment capability discovery.} 
    The internal mechanisms of large neural networks are not interpretable in causal or mechanistic terms. Model behaviour can change qualitatively with scale, prompting configuration or fine-tuning. Safety arguments that rely on deterministic traceability or component-level understanding cannot be fully applied.
    Unexpected behaviours, such as prompt injection or role-play exploitation, emerge only after deployment-like testing~\cite{buhl2024safety}.
    \item \textbf{Evaluation without stable ground truth.}
    Many AI systems are evaluated against existing human or automated systems that were never formally certified. Hence, risk arguments become marginal: we infer safety by comparison ("no worse than baseline") or by proxy metrics such as expert consensus, simulated games, or risk scores~\cite{chen2025maria} 
    This weakens the deductive foundation of safety arguments and shifts the reasoning mode toward induction and statistical inference.
    \item \textbf{Model updates outside system owner control.}
    AI applications often depend on proprietary large language models maintained by third parties; model updates may occur outside the system owner's direct control, thereby undermining the assumption that the assured configuration can be frozen post-certification. Because these models change or update without notice, the nature and distribution of risks also evolve in ways users cannot anticipate or verify. Unlike aircraft software, which is frozen after certification, AI systems evolve continuously through fine-tuning, retraining, data refresh, model combination, or expanded tool access. These changes invalidate static evidence. Assurance must therefore be dynamic, maintaining traceability across versions and regenerating claims and evidence as the system evolves~\cite{carlan2024dynamic}.
    \item \textbf{Threshold-based acceptability.} Regulators and developers increasingly rely on quantitative thresholds, of compute use, model capability, or safety score, as proxies for acceptable safety~\cite{oecd2025thresholds} 
    In AI systems, however, these thresholds are often incomplete, overlapping, or context-dependent. Safety cases must accommodate multiple, sometimes inconsistent, decision criteria rather than a single binary pass/fail condition.
    

    \item \textbf{Capability emergence decoupled from specification.} AI capabilities can emerge with scale, fine-tuning, or system-level scaffolding in ways that are not explicitly specified in requirements. As a result, safety-relevant behaviours may not be fully traceable from the intended functionality to the design artefacts, thereby weakening the assumption that assurance can be achieved through specification-driven decomposition alone. 

    \item \textbf{Tight coupling between model behaviour and context engineering.} Model behaviour is strongly shaped by context engineering choices~\cite{hua2025contextengineering2}, including prompting strategies, retrieval augmentation, tool interfaces, and user workflows. This means that safety properties cannot be assessed independently of the surrounding system and its operational context, and assurance must account for how changes in context can materially alter risks. 

\end{itemize}


AI systems are not unique because they are uncertain, but because the locus and timing of uncertainty is shifted, and because assurance must operate under continual epistemic incompleteness, not just probabilistic risk. That framing avoids straw-manning traditional systems while sharpening the real distinction. Together, these properties mean that AI safety cases cannot rely on design-time determinism; instead, they must embrace evaluation-time discovery and run-time adaptation. 
Assurance becomes a process of continuous learning rather than static certification.

\textbf{Emerging practices in AI safety cases.}
Several efforts illustrate how the safety-case paradigm is beginning to evolve for AI. Buhl et al.~\cite{buhl2024safety} propose a structure that integrates traditional CAE patterns with AI-specific concepts, including model evaluation, interpretability analysis, and governance oversight. Their "frontier AI safety case" uses layered argumentation across technical, organisational, and societal domains.
C{\^a}rlan et al.~\cite{carlan2024dynamic} extends this thinking into dynamic safety cases, linking claims to live metrics through "Checkable Safety Arguments" and "Safety Performance Indicators" (SPIs). The key insight is that evidence must be continually refreshed as model behaviour and deployment contexts evolve. Similarly, Clymer et al.~\cite{clymer2025example} demonstrate how a safety case can be constructed for misuse safeguards in large language models, combining threat models, red-teaming, and organisational response plans as evidence.

These recent works suggest three broad shifts already underway:

\begin{itemize}
    \item From design justification to behavioural justification: Safety is demonstrated through what the system does, not only through what it was designed to do.
    \item From completeness to adaptability: The argument's credibility depends on its capacity to evolve and revalidate itself, not on being once exhaustive.
    \item From component-level reasoning to ecosystem reasoning: AI assurance must consider interactions between model, tools, and governance.
\end{itemize}

\textbf{Implications for reusable AI safety-case templates.}
The divergence between classical and AI safety cases motivates the need for reusable, AI-specific templates that encode appropriate logic, structure, and expectations for evidence. Such templates should enable practitioners to:

\begin{itemize}
    \item \textbf{formulate conditional claims} that explicitly capture context, control boundaries, and capability assumptions
    \item \textbf{structure arguments} that can combine deductive reasoning (for architecture and controls) with inductive and abductive reasoning (for evaluation and mitigation)
    \item \textbf{incorporate empirical, model-based, and governance evidence} in coherent patterns that can be updated dynamically.
\end{itemize}

\subsection{Positioning this study relative to existing AI safety case approaches}

Prior research on AI safety cases spans multiple directions, including governance-oriented proposals, domain-specific templates, dynamic management mechanisms, and worked case sketches. However, these approaches differ in scope, level of abstraction, and reusability. 
To clarify the distinct contribution of this study, we compare it against representative approaches across these categories.

Table~\ref{tab:positioning} positions our contribution relative to these approaches.
Our contribution unifies these strands into an integrated, structured foundation for designing AI safety cases, comprising a CAE taxonomy, reusable template meta-models, and a composable pattern library that can be instantiated and systematically maintained throughout the AI lifecycle.

\begin{table}[ht]
\footnotesize
\caption{Positioning of this study relative to representative AI safety-case approaches (illustrative, not exhaustive).}
\label{tab:positioning}

\begin{tabular}{p{0.10\textwidth}  p{0.25\textwidth}  p{0.25\textwidth}  p{0.30\textwidth}}
\hline
{Approach} & 
{Primary output} & 
{Scope and reusability} & 
{What this study adds} \\ \hline

Safety cases for frontier AI \cite{buhl2024safety,hilton2025scalable} &
Motivates safety cases for frontier AI governance; outlines practicalities of producing a frontier AI safety case &
Primarily high-level guidance for producing and using safety cases; does not provide an explicit and practical-level solutions &
Provides an \textbf{integrated CAE taxonomy, reusable template, and pattern} library that operationalises how to construct CAE chains across recurring AI assurance problems \\
Cyber inability safety-case template \cite{goemans2024template,goemans2024template} &
A concrete CAE template arguing a model lacks unacceptable offensive cyber capability; decomposes claims into sub-claims supported by evaluation evidence &
Single risk-focused template (cyber inability); provides strong “worked CAE template” but narrow coverage across assurance problem types &
\textbf{Generalises beyond a single template}: comprehensive taxonomy of AI claim/argument/evidence types and a pattern library spanning multiple assurance scenarios, including cyber, control, evolution, thresholds \\
Dynamic safety cases for frontier AI \cite{Erfan2020dynamic,Shreyas2022dynamic,Laure2025dynamic} &
Dynamic Safety Case Management System (DSCMS) concept for systematic, semi-automated safety-case revision over time; links claims to system state &
Lifecycle/update mechanism emphasised; demonstrated on a cyber-related template; less emphasis on a broad CAE taxonomy and reusable pattern design space &
Extends dynamic safety-case management by \textbf{providing a structured CAE design space} (taxonomy + reusable templates + patterns), clarifying which claim/argument/evidence types evolve and what evidence triggers updates \\
Sketch of an AI control safety case \cite{korbak2025sketch,clymer2024justifysafety} &
Worked “sketch” of a control safety case for LLM agents; highlights key claims and evidence (e.g., control evaluations) &
Case-study style sketch (control measures); illustrates claim/evidence dependencies but is not a general template library &
Elevates such \textbf{sketches into reusable patterns and templates} grounded in a compositional CAE taxonomy, with explicit reasoning logics and evidence families \\
\rowcolor{gray!10}This study &
\textbf{An integrated and structured CAE-based foundation} for designing AI safety cases, comprising a taxonomy, reusable template meta-models, and a composable pattern library; guidance for maintaining auditability under AI evolution &
Designed for \textbf{composability across AI system types and assurance contexts}; supports reuse, traceability, and update triggers &
\textbf{Unifies and systematises prior strands} by providing: (i) an integrated CAE taxonomy, (ii) reusable template meta-models, (iii) composable AI-specific patterns, and (iv) lifecycle-aware update mechanisms grounded in explicit reasoning logics \\
\hline

\end{tabular}

\end{table}

\section{Methodology} \label{sec:methodology}
\label{Research Methodoly}
\begin{figure}[ht]
  \centering
  \includegraphics[width=.7\textwidth]{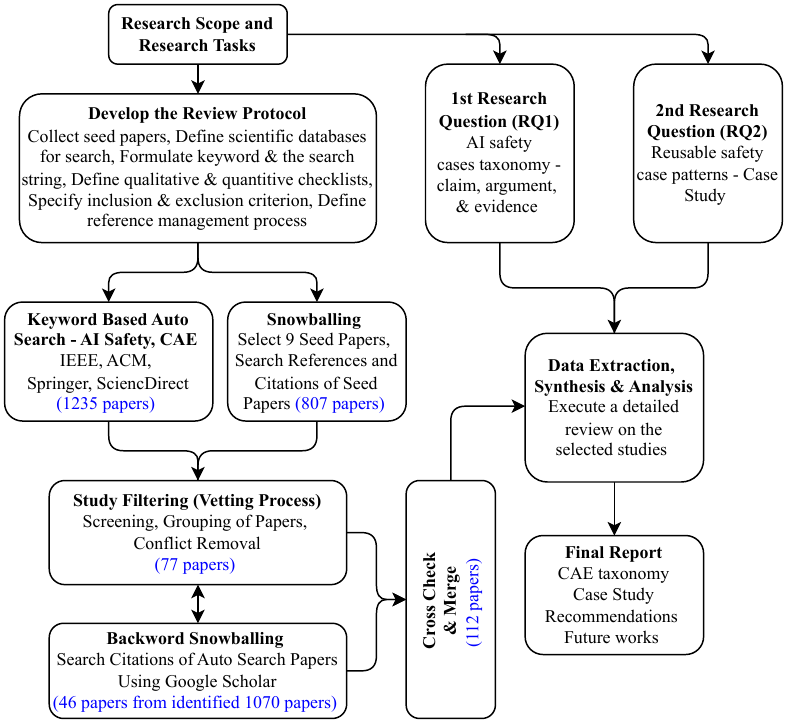}
  \caption[]{Architectural block diagram for this research paper}
  \label{WorkFlow_diagram}
  \vspace{-.2in}
\end{figure}

This study follows a Systematic Literature Review (SLR) approach based on established guidelines~\cite{kitchenham-guideline}. Figure~\ref{WorkFlow_diagram} presents the high-level workflow of this research. We first developed an SLR protocol that defined the research scope, search strategy, study selection criteria, and data analysis methods. In the following subsections, we describe each step in detail.

\subsection{Research Questions}
\label{Research Questions}
\textbf{P}opulation, \textbf{I}nterventions, \textbf{C}omparison, \textbf{O}utcomes, and \textbf{C}ontext (PICOC) can properly direct the formation of research questions for a review study~\cite{shamsujjoha_2021_SLR_MDD,shamsujjoha_ICSA-2025}. The PICOC for this study is shown in Table~\ref{PICOC-table}, following established guidelines~\cite{petticrew-systematic-picoc,kitchenham-guideline}.   
In this research, the \emph{Population} is the body of literature on AI safety cases, including CAE structures, safety case notations, and closely related assurance-case approaches. The \emph{Interventions} of interest are the methods, processes, templates, tools, and taxonomies used to construct, analyse, or operationalise AI safety cases. We consider an \emph{Outcome} successful if it contributes to reusable safety-case solutions, such as a CAE taxonomy and an associated pattern library. The relevant \emph{Context} is limited to works that explicitly connect AI systems with safety case development; this excludes generic AI ethics or governance works that lack a clear safety-case linkage. 
Our objective in this study is to analyse how safety cases are currently constructed for AI systems and to synthesise reusable CAE structures that support requirement-aligned, auditable safety cases. Within this scope, we formulated the following key research~questions:

\begin{description}
  \item[RQ1] \textbf{What taxonomy of claim types, argument strategies, and evidence categories is suitable for AI safety cases? }
  
  This research question aims to define classification schemes and templates for the kinds of claims (e.g., absolute vs. marginal safety claims), types of arguments (e.g., layered, barrier, or comparative arguments), and types of evidence (e.g., testing results, audits, formal proofs) that commonly occur in AI system assurance.
  
  \item[RQ2] \textbf{Can we identify reusable safety case templates and patterns for AI and demonstrate their use in a case study? }

Our second RQ plans to derive generic CAE templates and patterns for common AI safety concerns 
and test the study by applying the templates and a selected pattern in an illustrative case study of an AI system. 


\end{description}
\begin{table}[h]
\footnotesize
\caption{PICOC for this SLR}
\label{PICOC-table}

\begin{tabular}{p{0.15\textwidth}  p{.8\textwidth}}
\hline
{PICOC Element} & 
Description \\ \hline

{Population} & 
The literature on AI safety case, assurance case, claim argument evidence, goal structuring notation. \\ 
{Intervention} & 
Methods, processes, templates, tools, and taxonomies relevant to AI safety cases including  claim argument evidence centric approaches and frameworks.\\
{Comparison} & 
Exploratory synthesis across interventions. \\
{Outcomes} & 
CAE taxonomy, reusable pattern library, case study, application, evaluation criteria.\\

{Context} &
{Include:} Works that explicitly link AI safety case and CAE.
\newline
{Exclude:} General AI ethics and governance without safety-case linkage; security, privacy or access-control topics not framed as AI  or safety assurance; generic risk-management checklists; non-AI safety cases unless patterns are clearly transferable to AI; non-English sources; opinion pieces without evidence/method.
\\\hline
\end{tabular}
\vspace{-.2in}
\end{table}

\subsection{Search strategy}
\label{Search_Strategy}
We developed a strategy to search for papers that target AI safety and claim evidence and arguments to answer our RQs defined in Section~\ref{Research Questions}. The goal is to find as many primary study papers as possible. 
Our strategy consisted of four parts: search string identification, 
automatic search in an electronic database, snowballing using a seed paper, and backward snowballing using Google Scholar. With the assistance of the PICOC approach (Table~\ref{PICOC-table}), our search terms were divided into three primary concepts, as shown in Table~\ref{Concepts_Search_Terms}. These concepts helped us formulate a well-defined search string. 
\begin{table} [t]
\footnotesize
\caption{Concepts and search terms explanation}
\label{Concepts_Search_Terms}
\begin{tabular}{p{0.25\textwidth} p{0.7\textwidth}}

\hline
{Main Terms} & {Supportive Search Terms} \\ \hline
{Concept 1(Co1):} AI Safety & Safety case, assurance case, reliability, trustworthiness, acceptable risk, risk management, robustness, transparency, verification, validation, compliance etc.  \\  

{Concept 2 (Co2):} CAE and Structured Argumentation & Claim, argument, evidence, argument pattern, goal structuring Notation (GSN), pattern/module, assurance pattern, template etc. \\

{Concept 3 (Co3):} AI and related  domains & Autonomous system, intelligent system, Software, agents/multi-agent system. \\
\hline
\end{tabular}

\end{table}

We used alternative spellings, abbreviations, and synonyms of search terms to increase the number of relevant research papers. We used truncation and wildcard operators in finding these alternative keywords. Moreover, additional key terms or phrases identified during search iterations were added to the supporting search terms list. We assume they will collect all relevant articles. When constructing the final search query, the identified keywords, their alternatives, and related terms were linked with Boolean AND (\&\&) and OR ($\|$) operators. The OR operator was used to concatenate synonyms, and the AND operator was used to concatenate major concepts. A generic version of the search string we used is as follows:

\begin{tcolorbox}[breakable,colback=black!4,colframe=black!60,title= Generic Search String]
\begin{verbatim}
     (`safety' OR `safety case' OR `reliability' OR `trust*' OR `AI Safety')
AND (`claim-argument-evidence' OR `CAE' OR `structured argument*' OR `safety 
     justification' OR `evidence-based safety' OR `safety argument*' OR 
    `goal structuring notation' OR `GSN')
AND (`artificial intelligence' OR AI OR `machine learning' OR `autonomous 
     system*' OR `intelligent system*' OR `software')
\end{verbatim}
\end{tcolorbox}

\subsection{Study collection and filtering 
} 
\label{Collection-Filteration}
\begin{figure}[!ht]
  \centering
\hspace*{-.5in}\includegraphics[width=1.15\textwidth]{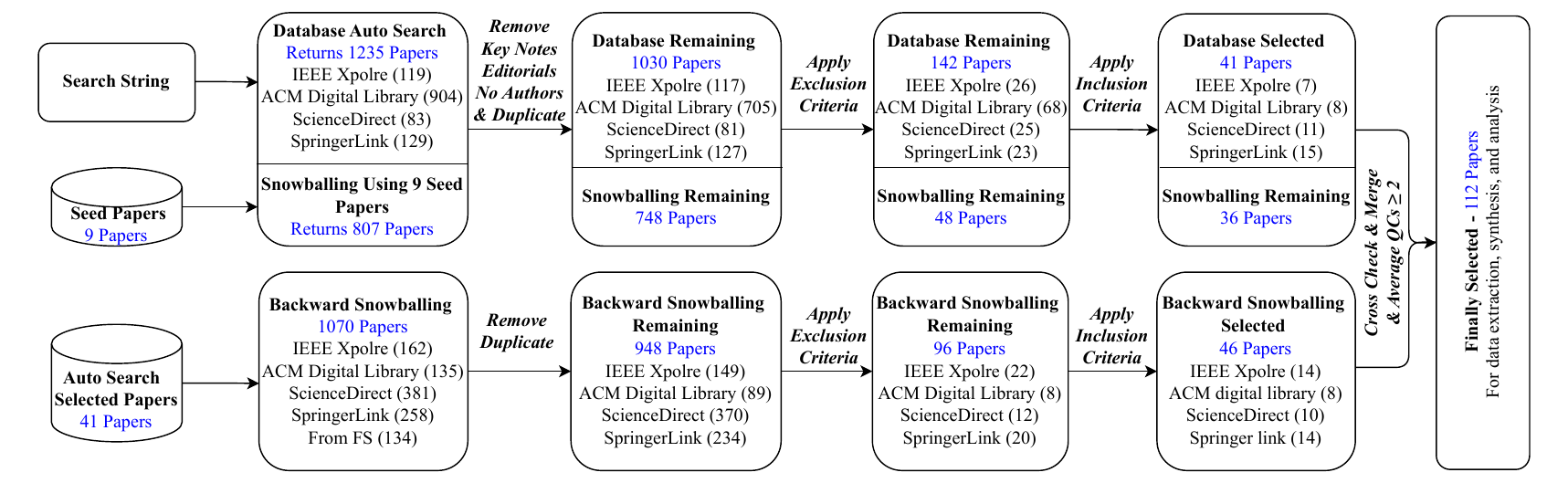}
    \caption[Primary study selection process with steps]{Primary study selection process steps} 
  \label{StudyFilteration}
\end{figure}

Our study filtration process is summarized in Figure~\ref{StudyFilteration}.  
%
We first ran the formatted search query on four major digital libraries, which returned 1,235 research papers. 
In parallel, we selected nine seed papers on AI safety and conducted snowballing, yielding 807 papers. 
The key rationale for selecting these papers as seed papers is discussed in \hyperref[AppendixA]{Appendix A}. We then removed 264 papers (205 from the database search results and 59 from the snowballing results of the nine seed papers) because they were duplicate articles, editorials, keynotes, or papers with no listed authors. After reading the title, abstract, conclusion, and, when necessary, skimming the introduction, methodology, and results sections, we applied \textbf{E}xclusion \textbf{C}riteria defined in Table~\ref{Inclusion_Exclusion_Criteria-table}. This step removed 1,588 papers in total. 
In the third step of filtration, we applied the \textbf{I}nclusion \textbf{C}riteria (IC) shown in Table~\ref{Inclusion_Exclusion_Criteria-table} and removed a further 113 papers 
that did not meet ICs. 

\begin{table}[hb]
\footnotesize
\caption{Inclusion and exclusion criteria}
\label{Inclusion_Exclusion_Criteria-table}
\begin{tabular}{l p{.46\textwidth} l p{.40\textwidth}}
\hline
{ECs} & {Exclusion Criterion} & {ICs} & {Inclusion Criterion} \\ 
\hline

EC$_1$ & Papers about AI but not use safety or claim argument evidence as key point of concern. 
& IC$_1$ & \multirow{3}{.40\textwidth}{Studies that propose, specify, evaluate, or apply safety cases / structured arguments for AI/ML systems or that provide foundational structured-argument constructs that are extracted to build the CAE taxonomy and template framework.} \\

EC$_2$ & Papers discussing AI or machine learning but not regarding safety case or CAE. 
& & \\

EC$_3$ & Safety papers without relevant AI/ML and without safety-case/structured-argument constructs beyond the scope. 
& & \\

EC$_4$ & Papers with inadequate information to extract (irrelevant papers). 
& IC$_2$ & \multirow{2}{.40\textwidth}{Content includes sufficient methodological or technical detail to extract and explicit linkage between AI risks or regulatory requirements and the safety case structure.} \\

EC$_5$ & Gray literature, workshop articles, posters, books, work-in-progress proposals, keynotes, editorials, secondary or review studies, vision papers with no concrete implementation. 
& & \\

EC$_6$ & Discussion papers and opinion papers, as well as surveys that do not include any solution. 
& IC$_3$ & Published or publicly available up to the review’s search date (August 2025). \\

EC$_7$ & Short papers less than three pages, irrelevant and low-quality studies that do not contain a considerable amount of extractable information. 
& IC$_4$ & Full-text conference papers, journal articles, regulatory documents, and credible industry publications that comply with the three concepts defined in Table~\ref{Concepts_Search_Terms}. \\

EC$_8$ & Conference or workshop papers if an extended journal version of the same paper exists. 
& IC$_5$ & Entire papers are written in English and use references. \\

EC$_9$ & Non-primary studies (secondary or tertiary studies). 
& IC$_6$ & Papers available in electronic format (e.g., doc, docx, pdf, HTML, ps). \\

\hline
\end{tabular}
\end{table}

We then carried out backward snowballing in Google Scholar using the 41 papers selected from the database search results. This yielded a further 1,070 papers. We followed the same process for this new set of papers and selected an additional 46 papers for inclusion that satisfied all ICs and met none of the ECs. At each step, at least two authors independently cross-checked the selection decisions and resolved any disagreements through discussion. Finally, after quality assessment (discussed in Section~\ref{Quality_Assessment}), we cross-checked and merged the initially selected sets, resulting in 112 primary studies for data extraction, synthesis, and analysis. These 112 original articles are summarized in \hyperref[AppendixB]{Appendix B}\footnote{We collected the paper list in August 2025; therefore, studies published after this date are not included in this review.}. Among the 112 selected studies, several (e.g., [S27], [S43], [S58], [S70]) do not explicitly target AI systems but were retained because they articulate foundational principles of structured, goal-based safety argumentation. Although not recent, these works continue to underpin contemporary assurance practices through their treatment of argument decomposition and evidence traceability. Moreover, their methodological contributions are domain-independent and directly inform the CAE-based taxonomy and reusable templates developed in this paper.

\subsection{Quality Assessment}
\label{Quality_Assessment}
We used a 1-to-5 numeric score (Very Poor, Inadequate, Moderate, Good, and Excellent) for \textbf{Q}uality \textbf{C}hecking (QC), applied to each study using the following eleven questions (QC$_1$ to QC$_{11}$). The questions focus on the relevance of the work to AI safety cases and to structured arguments, such as the claim-argument-evidence structure. 

\begin{description}
    \item[QC$_1$:] Is the study highly relevant to the research objectives and to the concepts of AI safety cases and CAE.

    \item[QC$_2$:] Does the study clearly explain the methodology that accomplishes its goals?

    \item[QC$_3$:] Does the study provide sufficient information on data collection, prototyping, and/or algorithms used?

    \item[QC$_4$:] Does the study present safety cases and a closely related structure for AI systems?

    \item[QC$_5$:] How well does the study detail how it has validated or evaluated its results?

    \item[QC$_6$:] Are clear outcomes and a corresponding results analysis reported?

    \item[QC$_7$:] Are the study limitations and possible future work adequately described?

    \item[QC$_8$:] What is the quality of the venue in which the study was published?

    \item[QC$_9$:] Are the implications and significance of the research findings discussed in the study?

    \item[QC$_{10}$:] Are the study's claim, presentation, and findings clear and understandable?

    \item[QC$_{11}$:] To what extent is the work presented in the study practically usable?

\end{description}

Each paper received an average score across these criteria. We set a threshold: if a study’s average \textbf{Q}uality \textbf{C}hecking (QC) score was below 2.0, we excluded it from our primary set as low-quality (11 studies from the selected list of 123 papers after cross-checking). Studies with borderline scores were discussed among the team, using the detailed extracted data to decide if they should be retained. 
As a result, the final analysis is based on a curated collection of high-quality publications that collectively address our research questions. This strengthens the validity of the conclusions and the proposed templates derived from the literature. QC score for all the selected studies is presented in \hyperref[AppendixC]{Appendix C}. In the selected primary studies, QC scores ranged from 2.73 to 4.55, with a mean of 3.37 on the 1-5 scale. Of the 112 studies in the final set, 96 ($\approx 86\%$) achieved a QC score of at least 3.0, including 22 ($\approx 20\%$) that scored above 4.0, while 16 ($\approx 14\%$) fell between 2.0 and 3.0 but were retained following team discussion of their relevance.

\subsection{Data synthesis and analysis} 
\label{Data Synthesis}
We first extracted all relevant information, including the RQ-specific items related to CAE structures and any proposed safety-case templates or patterns. Based on these data, we conducted a mixed-methods synthesis aligned with our two research questions (RQ1 and RQ2). On the quantitative side, we used simple descriptive statistics to identify patterns and trends across the selected studies on AI safety cases and structured argumentation. 


In addition to descriptive distributions, we conducted an explicit synthesis to characterize the state of the art in AI safety-case construction. This synthesis identifies (i) recurring strengths and weaknesses in how claims are formulated, e.g., conditional vs. unconditional safety, (ii) how arguments establish acceptability, e.g., reliance on comparators without explicit thresholds, (iii)   how evidence is operationalized, e.g., evaluation artefacts without stated quality criteria or traceability. We then map these findings to reusable template requirements, including fields, constraints, and update triggers, so that the study directly addresses observed gaps. 

We also conducted a thematic synthesis of the extracted CAE-related data. We grouped similar concepts and approaches across studies to derive higher-level themes concerning (i) how prior work classifies or organizes claims, arguments, and evidence for AI systems, leading to a structured CAE taxonomy and template for AI safety cases, (ii) how CAE elements are combined in practice e.g., common argument structures or evidence bundles that can be expressed as reusable safety-case templates or patterns for AI systems, and (iii) observations such as prevalent reliance on particular argument styles, dominance of certain evidence types, gaps in post-deployment or uncertainty-focused evidence, and examples of innovative or effective CAE practices. 
The derived themes were iteratively refined against the primary studies to ensure that the resulting CAE taxonomy, along with its associated templates and patterns, accurately reflected the underlying evidence. 

\begin{table} [!ht]
 \footnotesize
 \caption{Summary: Mapping of primary studies to CAE taxonomy categories. 
    }
    \label{taxonomy-mapping-table}
     \rowcolors{2}{gray!10}{white}
    \hspace*{-.5in}\begin{tabular}{p{0.065\textwidth} p{0.4\textwidth} p{0.4\textwidth} p{0.15\textwidth}}
    \hline
    \rowcolor{gray!15}
   Category & AI safety element: definition &  Core references & Inferred mapping\\ 
    \hline
         Claim & \textit{Assertion-based:} Claims that assert safety directly, either absolutely or relative to a comparator, without conditioning on specific constraints. &  S2, S7, S8, S10, S14, S15-S18, S20, S21, S22, S26, S33, S35, S41, S43, S45, S47, S49, S53, S59, S63, S70, S71, S74, S76, S81, S83, S88, S89, S91, S99 & S1, S3, S5, S30, S31, S107\\ 
         
         & \textit{Constrained-based:} Claims that safety holds only within defined operating boundaries such as context, modality, data regime, or validated envelopes. & S8, S10, S14, S15, S16, S17, S20, S23, S26, S28, S33, S35, S41, S43, S45, S47, S49, S53, S81, S83, S87, S88, S89, S93, S96, S112 & S1, S5, S18, S24 \\ 
         
         & \textit{Capability-based:} Claims that safety is ensured because the system’s abilities are restricted by design or intrinsic model behaviours (e.g., tool-access limits, refusal policies). & S10, S15, S26, S45, S47, S49, S53, S64, S81, S85, S86, S87, S89, S91, S93, S96,  \cite{amodei2016concrete,brundage2020toward,babcock2016agi}  & S17, S19, S28, S29, S78
         \\ 
          
         Argument & \textit{Demonstrative:} Arguments showing that layered controls, safeguards, or architectural mechanisms collectively satisfy safety criteria through deductive reasoning. & S2, S3, S5, S8, S13, S16, S17, S20, S21, S23, S26, S33, S35, S36, S44-S47, S49, S58, S63, S68, S69, S70, S71, S74, S81, S87, S88, S93, S96, S99, S100, S104, S106, S107, S108, S112 & S10, S4, S30 \\ 
         
         &\textit{ Comparative:} Arguments inferring safety by demonstrating that the system is no worse than a baseline or comparator across relevant metrics. & S8, S10, S16, S22, S33, S59, S81, S88, S89 & S18, S27, S65-67\\ 
         
         & \textit{Risk-based:} Arguments linking safety to formal risk estimation, uncertainty quantification, or compliance with risk-management frameworks. & S7, S8, S16, S21, S25, S30, S47, S63, S69, S70, S81, S83, S92, S93, S96, S99, S106 & S4, S18, S19,  S72, S107, S112\\ 
         
         & \textit{Casual and explanatory:} Arguments justifying safety by identifying, explaining, and mitigating underlying causes of failures or hazardous behaviours. & S26, S28, S47, S70, S71, S74, S76, S81, S83, S87, S93, S96, S100,  \cite{hawkins2010new,leveson2016engineering, barredo2019explainable, salay2019safety} & S24, S25, S73, S102  \\ 
         
         & \textit{Capability-oriented:} Arguments showing that safety is ensured through system containment, enforced limitations, or inherent model inability/refusal behaviours. & S8, S10, S47, S49, S64, S81, S84-S87, S89, S91 &  S111 \\ 
         
         & \textit{Normative and conformance:} Arguments deriving safety from adherence to recognised standards, guidelines, or organisational dependability principles. &  S6-S9, S25, S26, S28, S30,  S69, S70, S71, S74, S76, S81, S83, S87, S92, S93, S96, S99, S100, S106  & S11–S14, S16, S17, S20–S22, S32–S42, S45–S55, S57–S64, S105–S110 \\

         Evidence & \textit{Empirical:} Behavioural evidence obtained through testing, red-teaming, user studies, or evaluation under controlled and stress conditions. & S1, S2, S4, S8, S10, S13-S18, S20-S22, S26, S28, S33, S35, S41, S43, S44, S45, S47, S48, S49, S53, S55, S59, S70, S71, S74, S76, S77, S80, S81, S83, S85, S89, S93, S96, S98, S100 & S27, S56, S94, S103\\ 
         
         & \textit{Comparative:} Evidence from benchmarking or controlled comparisons showing performance against human or system baselines. &  S16, S28, S33, S46, S59, S64, S81, S89 & S18, S27 \\ 
         
         & \textit{Model-based:} Quantitative risk-analysis outputs derived from probabilistic models, simulations, or uncertainty propagation. &  S1, S9, S17, S20, S33, S35, S36, S44, S47, S49, S53, S55, S75, S81, S82, S83, S87, S93, S96, S98  & S100, S112\\ 
         
         & \textit{Expert-derived:} Structured expert judgment, scenario analysis, or horizon-scanning exercises used where empirical evidence is incomplete. &  S8, S13, S16, S22, S23, S35, S43, S44, S47, S48, S53, S63, S79, S81, S85, S93, S96, S100 & S1, S19, S97, S111  \\ 
         
         & \textit{Formal methods:} Mathematically rigorous analyses such as model checking, theorem proving, or static analysis demonstrating safety properties. & S7, S17, S20, S44, S49, S52, S53, S71, S80, S83, S93, S96, S98, S112 & S15, S19, S43, S63\\ 
         
         & \textit{Operational and field data:} Evidence from real-world operation, including logs, drift reports, monitoring dashboards, and governance artefacts. & S8, S19, S25, S46, S47, S49, S53,  S70, S74, S77, S81, S83, S85, S93, S96, S99, S100 & S22, S27, S28, S28,  S64,   S95, S101  \\ 
         
         & \textit{Mechanistic:} Evidence explaining model behaviour through mechanistic analysis, attribution, feature tracing, or subsystem verification. & S22, S63, S71, S81, S87, S89 &  S16, S19, S90, S111\\  
 
    \hline
    \end{tabular}
  \vspace{-.3 in}
\end{table}

\subsection{Summary information}

\begin{figure} [ht]
    \centering
    \subfloat[]{\includegraphics[width=0.55\textwidth]{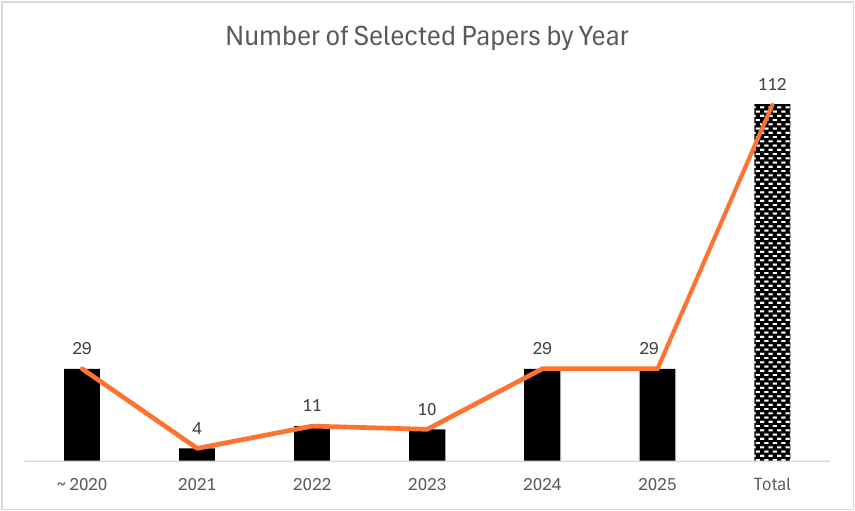}}
    \subfloat[]{\includegraphics[width=0.4\textwidth]{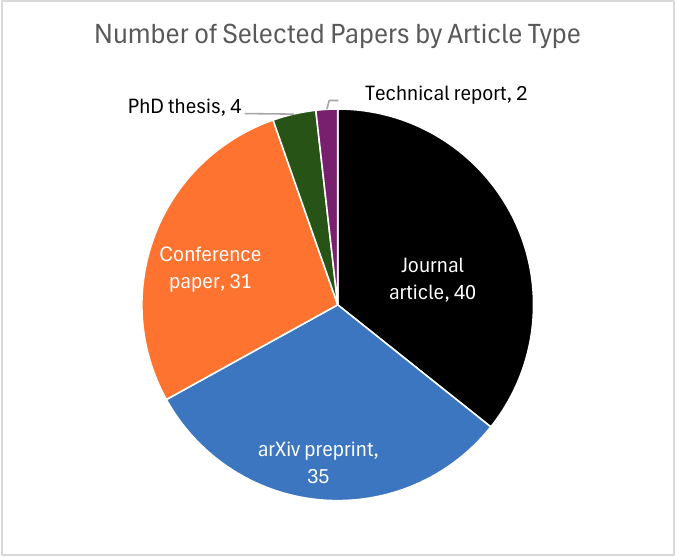}}
    \caption{The distribution of the selected articles in this study by published year and article types. (a) Number of selected papers by year; publications show a sharp increase in 2024 and 2025.(b) Number of selected papers by type; the majority are journal, conference, and recently posted arXiv papers.}
    \label{fig:number-yeartype}
\end{figure}

To provide an overview of the publication trends within our SLR dataset, Figure \ref{fig:number-yeartype} summarises the temporal and typological distribution of the 112 selected studies. 
Figure~\ref{fig:number-yeartype}(a) shows the number of selected papers by publication year; prior to 2020, the number of relevant works remains comparatively low and stable, with only modest growth between 2021 and 2023, but there is a sharp increase from 2024 onwards. Figure~\ref{fig:number-yeartype}(b) presents the distribution of article types and indicates that most studies appear as journal articles (40 out of 112, 35.71\%), arXiv preprints (35 out of 112, 31.25\%), and conference papers (31 out of 112, 27.67\%). 

\subsubsection{Overview of the selected studies}
Table~\ref{taxonomy-mapping-table} organises the CAE taxonomy into its Claim, Argument, and Evidence categories, each further divided into subcategories. For this mapping, in addition to the 112 papers selected through the SLR process, we also incorporated seven supplementary papers identified through a targeted search. These papers were not retrieved through the automated search or snowballing phases for two reasons. First, several of them use terminology that predates contemporary CAE- or AI-specific safety-case vocabulary, and second, some fall outside the scope of indexing by the chosen databases. Nevertheless, they represent influential or foundational works that provide important conceptual grounding for safety-case thinking, particularly regarding argument structures, inability-based reasoning, and safety-case reuse.

In Table~\ref{taxonomy-mapping-table}, the "Core reference" column records studies that explicitly instantiate a given subcategory in our analysis. For instance, studies S2, S7, and S8 state assertion-based claims as direct safety guarantees. Similarly, risk-based argumentation is represented where authors substantively engage hazards and mitigations, such as S18’s barrier-based reasoning to reduce exfiltration risk. The evidence layer captures explicit support. The empirical entries reflect reported evaluation artefacts such as red-team outcomes (e.g., S18 and S27). The formal-methods evidence is reserved for studies that provide proof-style assurance (e.g., S20) as part of a safety case. This treatment aligns with the presentation of defining categories with concrete, method-relevant examples.

The "Inferred mapping" column extends coverage by recording studies whose content functionally aligns with a subcategory even when the authors do not label it as such. This improves completeness but also introduces a methodological trade-off: implicit assignments can blur the boundary between the observed CAE framing and the inferred alignment. This increases susceptibility to interpretive bias unless the inference rules are explicit and consistently applied, as in S73 within causal and explanatory arguments. The following example illustrates both the benefit and the risk, considering  example studies:

\begin{tcolorbox}[colback=black!4, title= Example]
The mapping study S112, under a risk-based argument, highlights a substantive critique: safety cases should prioritize risk reduction over compliance. Similarly, adding S64 under operational/field evidence reflects an evidential stance that assumes the presence of runtime monitoring artifacts. However, these inferences should be read as analytical triangulation, not a re-labelling of author intent. We believe the table is strongest when implicit links are justified via clear criteria, e.g., required artifact types, explicit causal reasoning, or risk-management constructs, and reported as a validity mitigation rather than a substitute for explicit CAE usage in the studies.
\end{tcolorbox}


\subsubsection{Key insights into AI safety case construction}

Across the 112 analyzed studies, AI safety cases most commonly take the form of conditional claims supported by empirical evidence, often because ‘ground truth’ safety is unavailable in real deployments. However, we observe three recurring shortcomings.

\begin{itemize}
    \item \textbf{Acceptability criteria are underspecified}. Comparative or marginal claims frequently rely on a baseline (no worse than thresholds) without defining the comparator system’s assurance status or formalizing failure budgets and quantitative thresholds.

    \item \textbf{Evidence quality criteria and traceability are inconsistently stated.} While red-teaming, stress testing, and audit artefacts appear widely, many safety-case variants do not specify evidence independence, coverage, recency, reproducibility, or linkage, which limits auditability under continuous updates.

    \item \textbf{Post-deployment assurance remains fragmented.} A smaller subset of studies ($\leq 10\%$) link claims to operational monitoring and governance actions, which becomes increasingly necessary as models are updated, tool access changes, or deployment contexts drift.
\end{itemize}

Overall, these three major findings motivate our template requirements, i.e., each template instance must explicitly declare comparators/thresholds (when used), evidence quality criteria, and update triggers that require re-validation.

The reviewed literature indicates that AI safety case construction remains at a formative stage. While the CAE structure is widely adopted in form, its operationalisation varies significantly in claim definition, evidential adequacy, and lifecycle integration. The absence of reusable abstraction layers and formalised update logic limits composability and long-term auditability. This structural diagnosis motivates the need for systematic taxonomies and reusable template patterns to stabilise AI safety case engineering across domains.

\section{Reusable AI Safety-Case Templates} \label{sec:taxonomy}

\subsection{Rationale for reusable templates}
The emerging diversity of AI systems, from narrow-domain models to large-scale generative models, demands a systematic yet adaptable approach to structuring safety justification. In traditional engineering, safety cases are often bespoke, created for each project through expert judgement and iterative review. For AI, this approach does not scale. Model behaviours evolve too quickly, evidence sources are heterogeneous, and assurance must often be performed by teams without formal safety-case experience. 

A reusable, pattern-based approach can reduce subjectivity and promote consistency, while remaining flexible enough to handle uncertainty, discovery, and change.
Following \cite{bommasani2021opportunities} and adapting their template-based assurance logic for autonomous systems, the proposed solution defines reusable safety-case templates as standardised, semi-structured argument skeletons. Each template combines a claim pattern, a corresponding argument structure, and a set of evidence families appropriate to AI systems. These templates can be specialised or composed to construct end-to-end safety cases.

\subsection{Overview of the architecture} \label{sec:template}
The structure follows the classical CAE triad but adapts each element to AI-specific realities (Table \ref{tab:CAE architecture}).

\begin{table} [htb]
 \footnotesize
 \caption{Claim, Argument, and Evidence for AI systems}
    \label{tab:CAE architecture}
    \begin{tabular}{p{0.1\textwidth}p{0.31\textwidth}p{0.51\textwidth}}
    \hline
   Type & Purpose in traditional systems & Adaptation for AI systems\\ 
    \hline
         Claim & Fixed proposition of system safety (e.g., “System X is acceptably safe”) & Conditional and contextual proposition incorporating assumptions about data, model capability, and deployment context. \\

         Argument & Logical structure showing how subclaims and evidence support the top-level claim & Multi-modal reasoning combining deductive (design), inductive (empirical), abductive (mitigation), and statistical (uncertainty) inference \\

         Evidence & Test results, analyses, certifications & Empirical evaluations, red-teaming, interpretability analyses, governance artefacts, and risk-model outputs \\ 
    \hline
    \end{tabular}
    
\end{table}

The templates are designed to be composable. A single safety case may consist of multiple claim–argument–evidence chains stitched together; one addressing architectural robustness, another addressing misuse safeguards, and another addressing residual risk thresholds.

\begin{figure} [!ht]
    \centering
   \hspace*{-.2in} \includegraphics[width=1.05\textwidth]{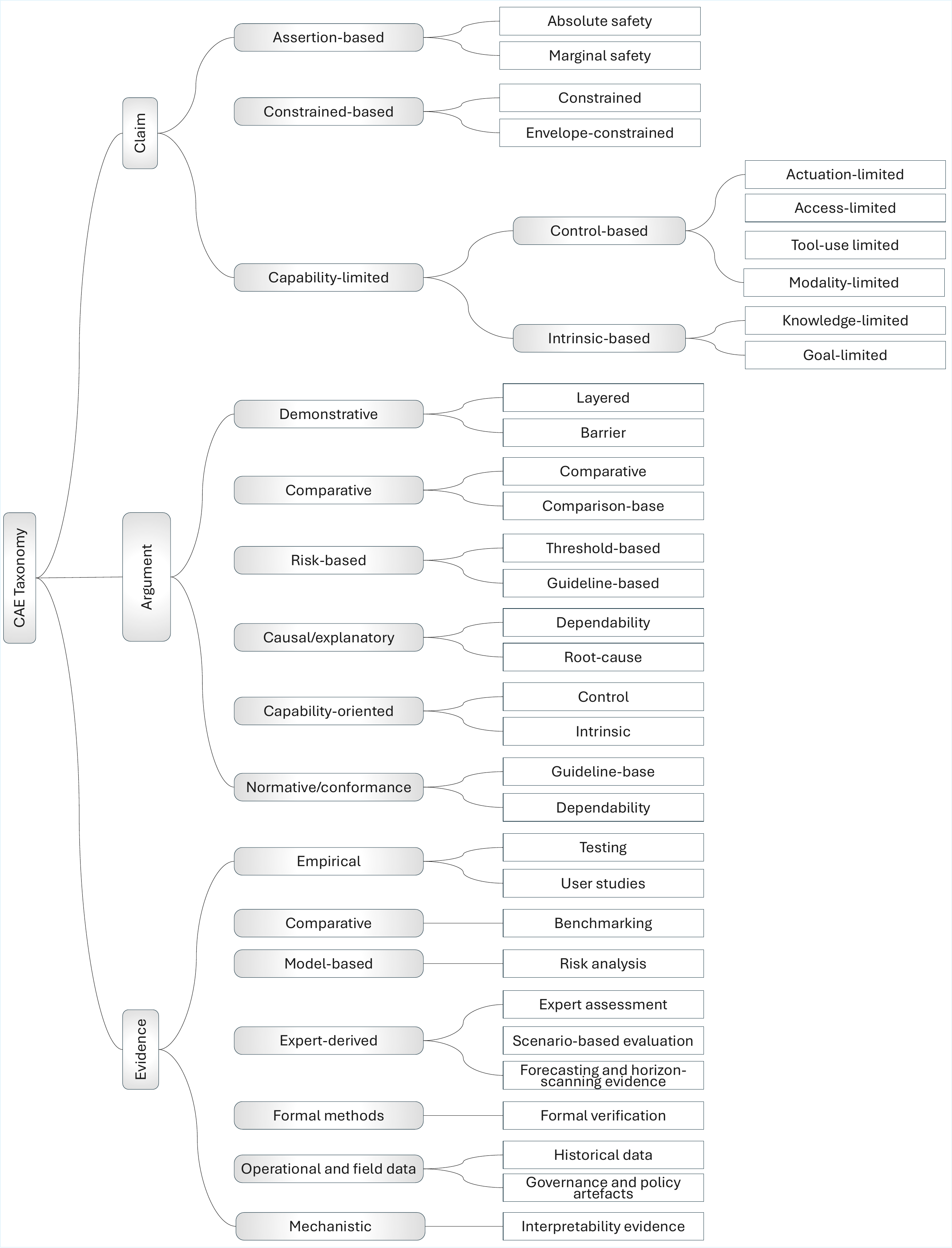}
    \caption{The CAE Taxonomy for AI systems.}
    \label{fig:CAE taxonomy}
    \vspace{-.2in}
\end{figure}

Figure \ref{fig:CAE taxonomy} shows the full CAE taxonomy for AI systems, which serves as the structural foundation for the proposed reusable safety-case template.
The taxonomy integrates AI-specific extensions into the classical CAE model, distinguishing how safety propositions (claims), reasoning structures (arguments), and supporting artefacts (evidence) interrelate across assurance layers.

\clearpage

The purpose of the taxonomy is to support structured composition and explicit reasoning, not to enforce classification purity or decision-tree logic. This distinction is critical for practitioners, as AI safety cases are inherently multi-modal and evolve through discovery, comparison, and governance over time.
Accordingly, the taxonomy is intentionally non-exclusive: certain claim, argument, or evidence types (e.g., guideline-based and dependability arguments) may appear under multiple categories. This reflects their dual role as both technical assurance mechanisms and normative proxies for acceptability, depending on the assurance intent and context of use.

\subsection{Claim} \label{sec:claim}
Claims in AI safety cases differ from those in conventional engineering because they must describe conditional and discovery-dependent safety propositions rather than absolute guarantees. Each claim type captures how safety is defined, bounded, or inferred within an evolving, data-driven system.
As shown in Figure \ref{fig:CAE taxonomy}, the taxonomy of claim categories is grouped by their assurance intent: context-based, constrained, capability-limited, and threshold-based.

\textbf{Assertion-based}

Assertion-based claims describe the mode of assurance, how safety is asserted relative to defined expectations or baselines. They include both absolute propositions (claiming acceptable safety outright) and comparative propositions (claiming safety relative to a reference system).

\begin{itemize}
    \item \textbf{Absolute safety claim:} A high-level claim that the system is acceptably safe for a specified purpose. This acts as a placeholder for more specific subclaims.\\
    Example: \textit{“AI System X is acceptably safe for document summarisation within approved enterprise environments.”}
    \item \textbf{Marginal safety claim:} Used when absolute safety cannot be quantified but relative safety can be inferred.\\
    Example: \textit{“AI System X is at least as safe as the previous deployed version Y (or a comparator Z) on relevant misuse benchmarks and consistency metrics”}
\end{itemize}

\textbf{Constrained-based}

Constrained claims state that safety is conditional on defined operational limits, such as data regimes, modalities, or environmental boundaries. They ensure that safety justification remains valid only within those bounds.

\begin{itemize}
    \item \textbf{Constrained claim:}
    Expresses that safety is bounded by particular operational conditions, data regimes, or user contexts. Context and constraints are treated jointly, as they are inseparable in AI (e.g., domain, modality, data source, tool access limits).\\
    Example: \textit{“AI System X is safe when interacting with text-only inputs drawn from internal data sources and used by authorised employees.”}
    
    \item \textbf{Envelope-constrained claim:} 
    AI safety is maintained only while the system operates within a validated performance or failure envelope, with mechanisms to detect and respond to out-of-envelope behaviour. Such claims emphasise runtime assurance, where the system transitions to a safe or degraded state when its operational limits are exceeded. \\
    Example: \textit{“System X is safe so long as it operates within a validated failure envelope, and transitions to a safe state when out-of-envelope behaviour is detected (e.g., by halting, degrading, or handing off to a human).”}
\end{itemize}

\textbf{Capability-limited}

This type of claim states that safety depends on capability boundaries, either \textit{control-based} (design choices, such as tool or network disablement) or \textit{intrinsic} (assumed AI system inability or refusal behaviour). 

\begin{itemize}
    \item \textbf{Control-based claim} depends on external or architectural design choices that enforce operational containment. This claim includes:
    
    \begin{itemize}
        \item \textbf{Actuation-limited} 
        aligns with prior safety research advocating containment and human-in-command architectures
        , which restricts AI systems from issuing direct actuation commands to prevent uncontrolled or unsafe physical actions. \\
        Example: \textit{“System X cannot issue direct actuation commands to industrial equipment.”}
        \item \textbf{Access-limited} asserts that the system cannot access unauthorised networks, data repositories, or sensitive resources. It limits data exfiltration and privacy risks.\\
        Example: \textit{“System X cannot connect to external databases or internet endpoints.”}
        \item \textbf{Tool-use limited} specifies that the AI may invoke only pre-approved or sandboxed tools, ensuring controlled interactions with external functions or APIs.\\
        Example: \textit{“System X can execute only whitelisted tools within its isolated environment.”}
        \item \textbf{Modality-limited} declares that the system is restricted to particular input or output modalities, preventing cross-modal risks such as audio or image manipulation.\\
        Example: \textit{“System X processes text only and cannot generate or interpret images, audio, or video.”}
    \end{itemize}

    \item \textbf{Intrinsic-based claim} depends on internal properties or alignment behaviours embedded in the model that make unsafe actions inherently infeasible. This means that refusal emerges from reinforced policies or internal gating rather than from external rule enforcement. If such refusal is implemented through wrappers, filters, or middleware, it constitutes a control-based claim.
    
    \begin{itemize} 
        \item \textbf{Knowledge-limited} states that the model cannot access or infer information beyond its authorised or trained data domain. It limits knowledge leakage or speculative inference about restricted topics.\\
        Example: \textit{“System X cannot retrieve or deduce information outside the approved knowledge base.”}
        \item \textbf{Goal-limited} 
        asserts that the system lacks persistent goals, self-modification ability, or independent task initiation beyond a defined session scope. This prevents uncontrolled decision loops or goal drift.\\
        Example: \textit{“System X cannot initiate tasks or alter parameters without explicit human instruction.”}
    \end{itemize}
    
\end{itemize}


Each claim type implies specific expectations for argument structure and evidence. For instance, contextual claims require evidence of boundary enforcement, whereas capability-limited claims require inspection of mechanisms or adversarial validation. 

\subsection{Argument} \label{sec:argument}

Arguments provide the connective reasoning that links claims to supporting evidence. AI safety cases require argument structures that can integrate heterogeneous reasoning modes and adapt as new evidence emerges. The proposed taxonomy defines two orthogonal dimensions, including function and logical form, which can be combined modularly.

\subsubsection{Argument taxonomy}

\textbf{Demonstrative} 

Demonstrative arguments show that all architectural and procedural layers collectively satisfy acceptance criteria.
They are used when safety depends on the completeness and integrity of design-time controls.

\begin{itemize}
    \item \textbf{Layered safety argument} decomposes a top-level claim into sub-claims covering independent assurance layers (e.g., data, model, governance).\\
        Example: \textit{“Each layer (e.g., data validation, model robustness, and human oversight) independently satisfies its acceptance criteria, and together they ensure overall system safety.”}   
    \item \textbf{Barrier argument} demonstrates that multiple independent safeguards exist to prevent or mitigate harm, ensuring redundancy.\\
        Example: \textit{“If automated content filters fail, human review and escalation protocols provide a secondary safety barrier."}
\end{itemize}
Demonstrative arguments primarily employ deductive reasoning and rely on architectural evidence, safety design documentation, and compliance verification.

\textbf{Comparative}

Comparative arguments justify safety by relative performance against a baseline or reference system.
They are central to marginal-risk and model-update cases.

\begin{itemize}
    \item \textbf{Comparative argument} establishes that the target system’s behaviour is no worse than a comparator across defined safety dimensions.\\
    Example: \textit{“System X’s misclassification rate is not higher than that of the human baseline at 95 \% confidence.”}
    \item \textbf{Comparison-based argument} extends comparison to include context alignment and fairness of reference—showing that differences in scope or data do not bias results.\\
    Example: \textit{“System X is similar to System Y, which is safe → therefore X is safe."}
\end{itemize}
Comparative arguments use inductive, statistical, and abductive reasoning. They are supported by benchmarking, controlled evaluations, and evidence from expert reviews.

\textbf{Risk-based} 

Risk-based arguments justify safety through explicit reasoning about risk levels or conformance with risk-management frameworks.
They are used when assurance depends on either quantitative risk estimation or adherence to recognised safety guidelines.

\begin{itemize}
    \item \textbf{Threshold-based argument} 
    demonstrates that the system’s estimated residual risk is acceptable, based on formal analysis or statistical modelling. This form directly connects quantitative risk evidence to the top-level safety claim.\\
    Example: \textit{“According to our risk assessment, System X does not pose unacceptable risk; our risk analysis demonstrates that risk levels are within tolerable bounds."}
    \item \textbf{Guideline-based argument} demonstrates that the system complies with an established safety or risk-management framework and infers acceptability from that compliance. It indirectly argues that adherence to recognised safety processes implies an acceptable level of risk.\\
    Example: \textit{“System X adheres to our safety framework; no system that follows this framework poses unacceptable risks.”}
\end{itemize}

While the typical rationale for a threshold-based argument is statistical reasoning, a guideline-based argument relies on deductive reasoning.

\textbf{Causal/explanatory} 

Causal or explanatory arguments link observed safety outcomes to identifiable mechanisms or mitigations.
They show that hazards are understood and effectively controlled.

\begin{itemize}
    \item \textbf{Dependability argument} demonstrates that system reliability and fault-tolerance mechanisms causally ensure safety. For example, it can incorporate reliability, availability, maintainability, and safety (RAMS). \\
    Example: \textit{“System X is dependably controllable: the architecture enforces fail-safe degradation and human handover on fault.”}
    \item \textbf{Root-cause argument} draws on causal-explanatory reasoning proposed in safety-case literature 
    and recent AI assurance studies 
    where safety is justified by identifying, explaining, and mitigating the underlying causes of observed failures.\\
    Example: \textit{“We justify the safety of System X by explaining and eliminating the previously observed risk. We observed [O1–O3] and evaluated competing hypotheses [H0…Hn] using [D1…Dk] (ablations, audits, counterfactuals). Mitigation M removes the hazard, and M was deployed."}
\end{itemize}

Causal arguments rely on abductive and deductive reasoning, drawing evidence from incident analyses, interpretability studies, and post-mitigation verification.

\textbf{Capability-oriented}

Capability-oriented arguments support capability-limited claims by showing that technical or intrinsic boundaries effectively constrain behaviour.

\begin{itemize}
    \item \textbf{Control argument} demonstrates that design or configuration controls prevent the system from exceeding authorised capabilities.\\
    Example: \textit{“System X is capable of causing serious harm, but it is restricted from using those capabilities; The AI system will not be modified such that it can use its capabilities to cause serious harm; The AI system’s model weights will not be stolen."}
    \item \textbf{Intrinsic argument} 
    shows that internal model behaviours (e.g., refusal, bounded reasoning) inherently limit risk.\\
    Example: \textit{“System X is not capable of causing serious harm, regardless of how it is modified or used; Safety-alignment training enforces refusal to generate disallowed instructions across tested scenarios.”}
\end{itemize}

These arguments combine deductive reasoning (for design rules) and empirical reasoning (for behavioural validation) and are evidenced by mechanism inspection, alignment evaluation, and red-team testing.

\textbf{Normative/conformance} 

Normative/conformance arguments demonstrate adherence to established frameworks, governance standards, or dependability engineering principles widely recognised as proxies for safety assurance.
This argument type demonstrates that system safety is justified through compliance with normative expectations rather than new empirical discovery.
This argument includes two types: Guideline-based argument and Dependability argument.
Although these argument types also appear under other functional categories (e.g., risk-based or causal/explanatory), the taxonomy is not exclusive: certain reasoning forms can legitimately contribute to multiple assurance dimensions depending on the intent and context of their use.
In this category, the emphasis is on institutional acceptability and auditability rather than on demonstrating reductions in new technical risk. Compliance with recognised frameworks or dependability doctrines functions as a proxy for acceptable safety within established governance regimes.

\begin{itemize}
    \item \textbf{Guideline-based argument} demonstrates that the system complies with an established safety or risk-management framework and infers acceptability from that compliance. It indirectly argues that adherence to recognised safety processes implies an acceptable level of risk.\\
    Example: \textit{“System X adheres to our safety framework; no system that follows this framework poses unacceptable risks.”} 
    \item \textbf{Dependability argument} demonstrates that system reliability and fault-tolerance mechanisms causally ensure safety. For example, it can incorporate reliability, availability, maintainability, and safety (RAMS). \\
    Example: \textit{“System X is dependably controllable: the architecture enforces fail-safe degradation and human handover on fault.”}
\end{itemize}

\subsubsection{Reasoning logic}

A safety case is not simply a repository of evidence; it is an argument about why the evidence justifies trust in a system’s safety \cite{rushby2015interpretation}. For AI systems, the reasoning logic underpinning this argument must reflect both the probabilistic nature of learning models and the socio-technical uncertainties of deployment. Traditional engineering relies predominantly on deductive logic (“if all components are safe, the system is safe”). AI assurance, by contrast, requires a pluralistic combination of deductive, inductive, abductive, statistical, and analogical reasoning, each appropriate to different parts of the assurance problem.

The choice of reasoning logic determines how claims are substantiated, how confidence is accumulated, and how residual uncertainty is expressed. The following subsections describe when and how each reasoning form should be applied.

\textbf{Deductive- structural and rule-based assurance}


Deductive reasoning is used when system-level safety can be inferred from formally specified properties, architectures, or controls. 
Recent work has demonstrated how such deductive assurance can be formalised using safety contracts and logical composition, including approaches based on Subjective Logic to represent confidence while preserving deductive argument structure \cite{herd2024deductive}.

Its typical use includes: 

\begin{itemize}
    \item Proving containment and control effectiveness (e.g., “If the AI system has no external network 
    access, it cannot exfiltrate sensitive data”).
    \item Establishing logical completeness of safety arguments under declared assumptions.
    \item Demonstrating conformance with fixed regulatory or procedural rules.
\end{itemize}

It provides rigour and traceability and supports formal verification or policy-based reasoning \cite{basir2009deriving}. Yet, it requires stable premises and is rarely sufficient for emergent or probabilistic behaviours.

The example scenario is:

\textit{"A newly developed “AI-based document evaluation system” has four safety layers (policy, testing, controls, and monitoring). Each layer meets its defined acceptance criteria, and all interfaces between layers are verified.\\
For example,
\begin{itemize}
    \item \textbf{Policy layer:} The system has documented risk classification and approval before deployment.
    \item \textbf{Testing layer:} The system is classified as high-risk; according to the high-risk criteria, it passed fairness and robustness tests with $\geq$ 95\% success.
    \item \textbf{Control layer:} According to the control requirements, only approved users can modify model parameters (access control enforced).
    \item \textbf{Monitoring layer:} Confirmed that incidents or model drifts are detected and logged within 24 hours.
\end{itemize}
As the AI system passed the four layers, it’s acceptable and safe."}

\textbf{Inductive- empirical generalisation from evaluation}

Inductive reasoning generalises from observed evidence, such as evaluation results, to infer overall safety \cite{oztekin2010inductive}.

Its typical use includes: 
\begin{itemize}
    \item Aggregating results from test datasets, red-teaming, or user trials to estimate expected performance.
    \item Establishing non-inferiority to a comparator based on statistical sampling.
    \item Supporting discovery-driven safety claims where behaviour must be inferred from empirical regularities.
\end{itemize}

While this logic captures real-world performance and adapts to continual evaluation, confidence depends on representativeness and coverage of the test conditions, and induction cannot guarantee future behaviour \cite{leake1996case}.

There are two example scenarios.\\
\textit{"\textbf{Scenario 1:} System X is an updated LLM model of System Y and Z (previous version of the model), both of which have been used safely in production for over a year. Since the latest version of the model meets or exceeds their safety benchmarks, it is inferred that it is also safe."\\
"\textbf{Scenario 2:} System X is a new LLM. After comparison with widely used LLMs (e.g., ChatGPT and Gemini), its performance on safety benchmarks is equal to or better than that of these LLMs. Therefore, System X is considered safe and acceptable."}

\textbf{Abductive- explanation and mitigation of anomalies}

Abduction supports reasoning about why a system failed or behaved unexpectedly and whether the proposed mitigation plausibly prevents recurrence.

This logic is typically used for:
\begin{itemize}
    \item Analyze incidents or findings of the red-team to infer causal mechanisms (e.g., “the refusal failure is due to hidden prompt leakage”).
    \item Linking discovered hazards to mitigation strategies when causal mechanisms are only partially understood.
    \item Building evidence chains connecting interpretability results with system-level safety claims.
\end{itemize}

It enables learning-oriented assurance and transforms unexplained anomalies into explainable risks with documented hypotheses \cite{burton2022causal}. Its limitations include reliance on expert judgment and the plausibility of explanations, as well as susceptibility to confirmation bias without cross-validation.

The example scenario is: \\
\textit{"AI-based risk management system kept crashing after long use. Engineers hypothesised it was due to memory leaks (H1), corrupted user data (H2), and overheating (H3). After testing/observation, confirmed memory leaks were the root cause (i.e., updated the memory management module, and the crashes stopped)."}

\textbf{Statistical- quantifying uncertainty and marginal risk}

Statistical reasoning is central to quantifying residual uncertainty, estimating harm probabilities, and supporting threshold-based or marginal-risk claims \cite{burton2023addressing}.

This logic is primarily used for:
\begin{itemize}
    \item Non-inferiority testing in marginal safety comparisons.
    \item Confidence-bound estimation of harm rates, false refusals, or false acceptances.
    \item Risk aggregation and uncertainty propagation across heterogeneous metrics.
\end{itemize}

It provides quantitative credibility and clear confidence intervals, but is sensitive to metric design, data bias, and assumptions of independence or stationarity \cite{denney2011towards}.

The example scenario is:\\
\textit{"Process analysis AI system has been newly developed. The organisation’s risk threshold for the AI system is 0.25. Based on testing and field data, the system’s estimated risk score is 0.18, below the organisation’s approved threshold of 0.25. Since \textbf{risk score $\leq$ threshold}, the system is considered acceptable for deployment."}

\textbf{Analogical- leveraging similarity and prior assurance} 

Analogical reasoning justifies safety by reference to similar, previously validated systems, datasets, or operating environments \cite{porter2024principles}.

It is used for:
\begin{itemize}
    \item Transferring assurance arguments from a comparable AI system (e.g., “safety controls proven for System Z apply to System X with minor adaptation”).
    \item Reusing validated templates and patterns for related applications or system versions \cite{kelly1999arguing}.
    \item Supporting initial assurance of novel systems by analogy before sufficient empirical evidence exists.
\end{itemize}

It enables rapid safety case construction when data or evaluation results are limited.
Yet, it requires careful justification of similarity conditions; analogies degrade as systems diverge in architecture or context.

The example scenario is as follows.\\
\textit{"System X, a new AI-assisted diagnostic tool, uses the same model architecture and data validation pipeline as System Y, which has been safely deployed in hospitals for two years. Since only the interface has changed and all core safety features are identical, System X is inferred to be safe by analogy."}


\textbf{Integrative reasoning and meta-assurance}

No single reasoning mode suffices for AI assurance. Deductive reasoning establishes formal bounds, inductive and statistical reasoning quantify empirical confidence, abductive reasoning explains and mitigates emergent issues, and analogical reasoning bootstraps new cases from prior experience. Mature safety cases must explicitly declare which reasoning logics apply to which claim types and how transitions between them are validated, for example, from abductive hypotheses to inductive evidence once mitigations are tested.

Meta-assurance, reasoning about the quality of reasoning, becomes essential. It requires showing that each logic is used under its valid conditions, that assumptions are traceable, and that uncertainty is bounded both within and across reasoning modes. 
For example, each logical form includes an associated pass condition that defines its validity criteria. These pass conditions collectively enable meta-assurance by allowing the reasoning process itself to be verified for correctness, coherence, and justified application. 

\subsubsection{Alignment between argument types and reasoning logics}

Although integrative reasoning combines multiple logical forms to strengthen overall assurance, each argument type primarily relies on a specific reasoning logic that determines how evidence supports its claim.
Understanding this alignment clarifies when a given argument form is appropriate and how different logics (anological, deductive, inductive, abductive, or statistical) collectively strengthen assurance credibility.
Table \ref{tab:argument-reasoning} summarises the relationship between argument types, their dominant reasoning logic, and a concise description of their assurance role.

\begin{table} [htb]
 \footnotesize
  \caption{Summary of the argument types and reasoning logics mapping}
    \label{tab:argument-reasoning}
    \begin{tabular}{p{0.15\textwidth}p{0.18\textwidth}p{0.59\textwidth}}
    \hline
   Argument & Reasoning logic & Description\\ 
    \hline
    Layered safety & Deductive & Demonstrates that all assurance layers (e.g., data, model, governance) collectively satisfy safety criteria. \\
    Barrier & Deductive & Shows that multiple independent safeguards exist, ensuring redundancy even if one control fails.\\
    Comparative & Inductive or Statistical & Generalises from empirical comparison to infer that the AI system is no worse than a reference baseline.\\
    Comparison-based & Analogical & Justifies safety by reference to similar, previously validated systems, datasets, or operating environments. \\
    Threshold-based & Statistical & Quantifies residual risk and demonstrates that it remains below the defined numerical threshold.\\
    Guideline-based & Deductive & Infers safety from adherence to recognised frameworks, standards, or organisational policies.\\
    Dependability & Deductive or Statistical & Establishes that reliability and fault-tolerance mechanisms causally ensure acceptable safety levels.\\
    Root-cause & Abductuve & Provides causal explanations for observed failures and links corrective actions to prevention of recurrence.\\
    Control & Deductive & Demonstrates that external design or configuration controls effectively restrict system capabilities.\\
    Intrinsic & Deductive & Shows that built-in model properties (e.g., alignment or refusal behaviour) inherently prevent unsafe actions.\\
    \hline
    \end{tabular}
   
\end{table}

\subsection{Evidence} \label{sec:evidence}

Evidence substantiates the premises of arguments. 
In AI safety cases, credible evidence must capture both technical and governance realities, showing not only that the system behaves safely under test conditions but also that safety controls, policies, and monitoring processes remain effective over time. 
Unlike conventional engineering evidence, which is often deterministic and static, AI evidence must be empirical, dynamic, and socio-technical. It combines computational validation with organisational verification.

The taxonomy in Figure \ref{fig:CAE taxonomy} organises AI safety evidence into seven primary categories, including empirical, comparative, risk-based, expert-derived, formal methods, operational, and mechanistic. Each represents distinct modes of substantiation that map to different reasoning logics and argument functions introduced in Section \ref{sec:argument}.


\textbf{Empirical}

Empirical evidence establishes behavioural facts about the AI system through direct observation and evaluation.
This primarily includes testing and user studies, which can be used with red-teaming, fuzzing, or scenario simulations.

\begin{itemize}
    \item \textbf{Testing} includes results from functional, stress, adversarial, or red-team tests showing how the system behaves under controlled or extreme conditions.
    \item \textbf{User studies} refer to structured experiments involving human participants that assess usability, fairness perception, and effectiveness of oversight mechanisms.
\end{itemize}

The quality focus of empirical evidence is on the representativeness of datasets, the reproducibility of test procedures, and the coverage of operational contexts.

\textbf{Comparative benchmarking}

Comparative benchmarking evidence demonstrates relative safety through reference evaluations.
It strongly supports comparative arguments.
The evidence type includes benchmarking results, comparative performance on safety-related benchmarks, or competitions
are used as controlled studies comparing the AI system to human or legacy systems using equivalent metrics.

The quality focus includes the fairness of comparators, the equivalence of the evaluation scope, and the statistical significance of differences.

\textbf{Model-based risk analysis}

Model-based evidence provides quantitative reasoning about residual risk and uncertainty.
Potential risk models include Bayesian networks, causal models, or Monte Carlo estimations of harm likelihood.
While probabilistic or Bayesian risk analysis involves statistical estimation of harm probabilities and confidence intervals, simulation or Monte Carlo evaluation propagates uncertainty across system variables to determine threshold compliance.

The quality focus is on transparency of assumptions, calibration with empirical data, and sensitivity analysis of parameters.

\textbf{Expert-derived} 

Expert-derived evidence captures structured professional judgement where empirical or model-based data are incomplete.
It supports abductive and analogical reasoning.


\begin{itemize}
    \item \textbf{Expert assessment} is formal elicitation or review by qualified experts estimating the likelihood, severity, or adequacy of controls. It includes Delphi studies.
    \item \textbf{Scenario-based evaluation} includes tabletop or synthetic exercises exploring plausible failure modes and mitigation strategies.
    \item \textbf{Forecasting and horizon-scanning evidence} 
    is anticipatory analysis that identifies emergent capabilities, threat vectors, or long-term socio-technical impacts.
\end{itemize}

The key quality aspects include the diversity of expertise, the transparency of reasoning, and the explicit representation of uncertainty and consensus.

\textbf{Formal verification}

Formal verification provides mathematically rigorous assurance that specified safety or containment properties hold. Typical methods include model checking, theorem proving, static analysis, and symbolic execution to establish invariants, enforce rules, and verify bounded-autonomy constraints. 
It primarily supports deductive reasoning within demonstrative and capability-oriented arguments. 

The quality focus is on the soundness of proofs, tool qualification, and the correspondence between the verified model and the implemented system, with clear traceability of assumptions and coverage of critical properties.

\textbf{Operational and field data }

Operational evidence validates safety during real-world use.
It supports both demonstrative and statistical arguments by confirming that the implemented controls remain effective.

\begin{itemize}
    \item \textbf{Historical data} includes logs, drift reports, and post-deployment metrics confirming stability and safe performance.
    \item \textbf{Governance and policy artefacts} 
    refer to monitoring dashboards, audit trails, and approval records linking operational data to oversight actions.
\end{itemize}

The quality focus is on data recency, record authenticity, and traceability to system versions.

\textbf{Mechanistic or interpretability evidence}


Mechanistic evidence links internal model mechanisms to observable safety behaviour.
It strongly supports causal/explanatory and capability-oriented arguments.

The evidence type includes interpretability analysis, such as attribution mapping, feature tracing, or causal-path exploration, that explains why behaviour is safe or unsafe.
It also includes formal verification, such as proofs or static analyses, demonstrating that safety or containment properties hold within subsystems.

Evidence must also satisfy quality criteria such as independence, coverage, recency, reproducibility, and representativeness, enabling its credibility to be assessed systematically.

\subsection{Composability and reusability of the template} 
The true power of these templates lies in their composability. A full safety case can be built as a network of reusable modules.
Each module is reusable across systems and contexts, with only the specific claim instantiation and evidence artefacts substituted. The result is a structured but flexible assurance design; one capable of supporting AI systems whose safety can only be demonstrated through continual evaluation and rediscovery.

\section{Safety Case Patterns for AI Systems} \label{sec:pattern}


While Section \ref{sec:taxonomy} established the taxonomic foundation for constructing AI safety cases, practitioners ultimately need end-to-end patterns, reusable narrative structures that show how claims, arguments, and evidence interact to address recurring assurance challenges. A pattern is not a fixed template but a worked composition of the claim, argument, and evidence taxonomies tailored to a particular class of AI system risks. These patterns help align safety reasoning with how AI systems are actually developed and operated: iteratively, empirically, and under uncertainty.

Each pattern below is organised as follows:

\begin{enumerate}
    \item Problem or assurance challenge.
    \item Applicable claim types.
    \item Recommended argument functions and reasoning logics.
    \item Evidence families and quality conditions.
    \item Practical example structure.
\end{enumerate}

In this section, we introduce four representative patterns that correspond to the unique characteristics of modern AI assurance: "discovery-driven evaluation", "marginal risk without ground truth", "continuous system evolution", and "threshold-based acceptability decisions".
We also discuss integrating these patterns into composite safety cases used in practice.

\subsection{Discovery-driven evaluation pattern}

\textbf{Problem}

Traditional engineering safety cases rely on a full hazard enumeration prior to operation. For AI systems, this is infeasible: new capabilities and risks are discovered only after deployment-like testing. The assurance challenge is to justify safety in the face of incomplete knowledge and ongoing empirical discovery.

\textbf{Applicable claim types}
\begin{itemize}
    \item Contextual claims constraining safe operating boundaries (e.g., data domain, user group, or task category).
    \item Capability-limited claims, especially intrinsic limitations (e.g., refusal to execute unsafe content).
    \item Marginal safety claims compared to a reference system or control scenario.
\end{itemize}

\textbf{Argument functions and reasoning logics}

\begin{itemize}
    \item Demonstrative–Deductive for control enforcement (containment, isolation).
    \item Causal–Abductive to explain and mitigate newly discovered hazards.
    \item Risk-based–Statistical to express residual uncertainty and show progressive improvement.
\end{itemize}

\textbf{Evidence families}

Empirical testing, adversarial and stress evaluation, interpretability-based investigation of failures, and governance artefacts showing incident response or post-mitigation procedures. Evidence must emphasise recency, reproducibility, and completeness of evaluation coverage.

\textbf{Example structure}

\textbf{Top claim:} "AI system X is safe for interaction tasks in context C under control K."
\begin{itemize}
    \item Demonstrative argument: enforcement of access, isolation, and rate limits.
    \item Causal argument: identified risk R$_1$ (e.g., prompt leakage) explained via mechanism analysis and mitigated through instruction filtering M$_1$.
    \item Risk-based argument: post-mitigation evaluation shows risk rate r $\leq$ threshold t within statistical confidence interval.
    \item Continuous discovery process documented as part of the evidence base, with results linked to monitoring triggers.

\end{itemize}
    
This pattern formalises assurance through evaluation and discovery, embedding iterative empirical testing into the safety case rather than treating it as post-certification validation.

\subsection{Marginal-risk pattern without ground truth}

\textbf{Problem}

Many AI systems are deployed in domains where there is no fixed ground truth or where incumbent systems, whether human workflows, legacy automation, or peer AI systems, were never rigorously validated. In these cases, the objective is not to demonstrate absolute safety, but to show that the AI system is no less safe than an accepted comparator. Because risk must be inferred indirectly, assurance relies on a structured combination of predictability, capability, and interactive (game-based) metrics \cite{chen2025maria} that collectively approximate the system’s safety margin.

\textbf{Applicable claim types}
\begin{itemize}
    \item Marginal safety claim: “AI system X is at least as safe as comparator Z on relevant predictability, capability, and interaction metrics.”
\end{itemize}
    
\textbf{Argument functions and reasoning logics}

\begin{itemize}
    \item Comparative–Inductive to establish that AI system X matches or exceeds comparator Z on defined safety-related dimensions.
    \item Risk-based–Statistical to quantify non-inferiority margins and propagate uncertainty across heterogeneous metrics.
    \item Normative–Deductive to justify comparator selection and alignment with accepted operational or regulatory standards when empirical evidence is partial.
\end{itemize}

\textbf{Evidence families}

Evidence is drawn from multiple proxy domains to triangulate safety in the absence of a definitive benchmark:

\begin{itemize}
    \item Consistent predictability metrics – measures of consensus and stability that indicate behavioural coherence under small input perturbations or prompt reformulations. Examples include inter-annotator agreement with human experts, stability across rephrasings, and invariance under benign format changes.
    \item Capability-based metrics – controlled evaluations of the AI system’s competence in safety-relevant tasks, bounded by known capability envelopes. This may include stress tests on reasoning accuracy, constraint adherence, or refusal behaviour.
    \item Game-based metrics – structured interactive evaluations where human or AI agents probe the system in adversarial or cooperative scenarios to surface hidden failure modes. These “games” can simulate contest conditions (e.g., red-team versus defence) or cooperative judgment alignment tasks that reveal marginal safety performance under realistic pressure.
    \item Expert elicitation and structured judgement – panels assessing comparative harm likelihood and interpretability of results, supplemented by independent review of experimental fairness.
\end{itemize}

\textbf{Example structure}

\textbf{Top claim:} “AI system X is at least as safe as comparator Z for task Y under the evaluated proxy metrics.”

\begin{itemize}
    \item Comparative argument: define comparator Z and demonstrate alignment of task scope, data, and operational context; analyse differences and potential confounders.
    \item Predictability sub-claim: show that AI system X achieves equal or higher consistency, consensus, and stability scores than Z under controlled prompt or context perturbations.
    \item Capability sub-claim: present controlled task performance within safety-critical dimensions, demonstrating that observed capability boundaries remain within accepted operational limits.
    \item Game-based sub-claim: report results from adversarial or cooperative gameplay showing equivalent or reduced harm frequency and better recovery behaviour relative to Z.
    \item Risk-based argument: integrate the three metric families into a composite marginal-risk index; demonstrate non-inferiority $(\Delta \leq \delta)$ with statistical confidence $(\geq \alpha)$.
    \item Normative argument: justify why Z constitutes an appropriate reference system, referencing governance precedents or deployment track record.
    \item Evidence package: includes evaluation protocols, datasets, gameplay logs, aggregation scripts, expert-panel reports, and independent audits of experimental design and fairness.
\end{itemize}

This pattern operationalises marginal safety reasoning when ground truth is absent. It leverages multi-dimensional proxies such as predictability, capability, and interactive robustness to provide a defensible, quantitative, and qualitative basis for comparing an AI system’s safety against human or legacy baselines. It enables deployment decisions grounded in relative rather than absolute assurance, while preserving transparency in the reasoning chain and quality of evidence.

\subsection{Continuous-evolution pattern}

\textbf{Problem}
AI systems undergo continual modification, including model updates, retraining, data refreshes, scaffolding updates, and integration with new tools. Each change potentially alters risk characteristics, invalidating static evidence. The assurance challenge is to maintain a living safety case that stays valid through system evolution.

\textbf{Applicable claim types}

\begin{itemize}
    \item Contextual claims updated with versioned scope and control boundaries.
    \item Marginal safety claims comparing new and prior system configurations.
\end{itemize}

\textbf{Argument functions and reasoning logics}

\begin{itemize}
    \item Demonstrative–Deductive for update and rollback control mechanisms.
    \item Comparative–Inductive for version-to-version safety parity.
    \item Risk-based–Statistical for continuous performance metrics with alert thresholds.
\end{itemize}

\textbf{Evidence families}

Dynamic evaluation reports, regression test results, Safety Performance Indicators (SPIs) linked to claims, incident trend analysis, and governance logs for release approval. Evidence must include timestamps and automated traceability to prior versions.

\textbf{Example structure}

\textbf{Top claim:} “AI system X remains acceptably safe after update U.”

\begin{itemize}
    \item Demonstrative argument: update process verified by formal change-control procedure P.
    \item Comparative argument: no statistically significant degradation compared to prior version or alternative system Z on safety benchmarks.
    \item Risk-based argument: rolling SPI indicators within pre-declared acceptance bounds.
    \item Governance evidence: audit trail confirming revalidation and approval of updated safety case.
\end{itemize}

This pattern embodies the notion of a dynamic safety case, a living artefact refreshed alongside system development cycles.

\subsection{Threshold-comparator pattern}

\textbf{Problem}

Developers and regulators often define multiple quantitative thresholds (e.g., harm probability, compute expenditure, or autonomous capability) as proxies for safety acceptability. These thresholds may be incomplete or conflicting. The assurance challenge is to demonstrate that the AI system satisfies the relevant thresholds and justify how they are prioritised or combined.

\textbf{Applicable claim types}

\begin{itemize}
    \item Constraint-based risk claims with explicit numerical limits.
    \item Contextual claims clarifying conditions of applicability.
    \item Normative claims aligning thresholds with policy or external frameworks.
\end{itemize}

\textbf{Argument functions and reasoning logics}

\begin{itemize}
    \item Threshold-based, Deductive-Structural for mapping thresholds to system properties and decision rules.
    \item Statistical for evidence aggregation and uncertainty handling.
    \item Normative for explaining prioritisation when thresholds conflict or overlap.
\end{itemize}

\textbf{Evidence families}

Model-based risk analyses, empirical metrics, acceptance-criteria tables, sensitivity and worst-case analyses, and governance documentation describing threshold selection.

\textbf{Example structure}

\textbf{Top claim:} “AI system X meets defined safety thresholds T$_1$-T$_n$ for deployment in context C.”

\begin{itemize}
    \item Deductive argument: decomposition of system-level safety into measurable metrics tied to T$_1$-T$_n$.
    \item Statistical argument: aggregation function F(T$_1$-T$_n$) pre-declared, with uncertainty propagation and correlation analysis.
    \item Normative argument: justification for threshold choice hierarchy, consistent with policy or regulation R.
    \item Evidence: metric definitions, evaluation datasets, acceptance-criteria reports, and independent review summaries.
\end{itemize}

This pattern operationalises quantitative decision-making in safety cases, ensuring transparency about how thresholds translate into deployment approval or rejection.

\subsection{Integrating patterns into composite safety cases}

In practice, an AI system’s safety case will combine several of these patterns. For instance, a conversational agent might use the discovery-driven pattern to capture ongoing capability discovery and red-team evaluation, the marginal-risk pattern to justify deployment relative to human helpdesk baselines, and the continuous-evolution pattern to manage version updates. Each pattern contributes a module that can be updated independently while maintaining traceability to the overall safety claim.

Reusable patterns, therefore, enable composable assurance, the ability to build, maintain, and audit AI safety cases as evolving mosaics rather than static documents. This compositional approach allows alignment with both rapid development cycles and dynamic regulatory expectations.

\section{Case Study: AI-based Tender Evaluation System} \label{sec:case study}

\subsection{Case system overview and context}

Government tender evaluation processes traditionally involve two independent human reviewers who assess and score supplier submissions against defined criteria such as compliance, value for money, and risk. Discrepancies between reviewers are reconciled through discussion or third-party moderation.

The pilot AI-based tender evaluation system replaces one of the two human reviewers with an AI model.
The AI system was deliberately not trained on historical data. This design choice was critical to avoid bias and artificial correlation in the evaluation process. Training on historical samples would have inflated the “agreement” metrics, as outcomes would have been partially determined by prior examples rather than genuine reasoning. Instead, the system relied on a prompt-based approach using GPT-4o, in which both the document under evaluation and the criteria document (e.g., policy documents and scoring rubrics) were provided at inference time.
The system assists the remaining human reviewer by producing structured assessments and justifications. Its purpose is to improve efficiency and consistency while maintaining or enhancing fairness, transparency, and procedural integrity.

A key assurance challenge is the absence of an absolute ground truth: there are no definitive “correct” tender scores, only prior human judgements that vary across reviewers and contexts. Hence, safety and reliability must be justified marginally, so that the new AI + human configuration performs no worse (and ideally better) than the existing human + human process across comparable criteria.

This context makes the \textbf{Marginal-Risk Pattern without Ground Truth} the most appropriate safety-case pattern, as it supports reasoning about relative safety and integrity using proxy metrics and comparative evaluation.

\subsection{Applied pattern: Marginal-Risk Pattern without Ground Truth}

\textbf{Claim}

For this case, we selected "Marginal safety claim" and state:

\textit{"The AI-based tender evaluation process is at least as safe, fair, and reliable as the prior human-only process on relevant performance and integrity metrics."}

\textbf{Argument functions and reasoning}

To support the claim, practitioners can select argument types from three main argument functions (comparative, risk-based, and normative) depending on the availability of evidence and the decision context. These functions can be used individually or in combination to balance empirical and procedural assurance.

\begin{itemize}
    \item \textit{Comparative (Inductive reasoning)} is the primary argument function for marginal-risk cases where there is no ground truth but a reference system (e.g., human baseline) exists. It generalises from observed evidence to infer that the new system’s behaviour is no worse than the comparator across relevant dimensions. 
    This is recommended when direct empirical comparison is possible through shared metrics, datasets, or tasks.
    \item \textit{Risk-based (Threshold-based argument, Statistical reasoning)} is quantitative analysis estimates uncertainty and verifies non-inferiority $(\Delta \leq \delta)$ between configurations with confidence $(\geq \alpha)$, based on bootstrapped agreement and harm-probability measures. 
    This is recommended when the evaluation produces measurable indicators that can be statistically combined.
    \item \textit{Normative (Deductive reasoning)} supports assurance by referencing accepted practice, policy, or precedent, establishing that the comparator system or metric selection is valid. In this case, the comparator (human-human baseline) is justified as an appropriate reference, given regulatory and procedural acceptance in public-sector procurement.
\end{itemize}

These reasoning modes collectively support marginal assurance in the absence of ground truth and ensure that both empirical performance and normative alignment are represented.

For this case, two argument functions are applied: \textbf{Comparative and Risk-based (Threshold-based)}. 

The \textit{Comparative argument forms the primary reasoning mode} because the AI–human process is evaluated relative to an existing, operational human–human baseline. 

We defined \textbf{marginal risk} for this case: $MR = \Delta R = R_{\text{AI+Human}} - R_{\text{Human+Human}}$, 
where $R$ is a multi-dimensional risk vector encompassing performance, reliability, safety, security, fairness, privacy, compliance, cost, and resilience.

The \textit{Risk-based (Threshold-based) argument complements it} by quantifying residual uncertainty and demonstrating that performance differences fall within acceptable confidence bounds.

The Normative argument is not explicitly developed here, as the baseline comparator and procurement policies already define an accepted standard of practice.

Table \ref{tab:argument_selection} summarises the selected argument types and rationale of the selection for this case study.

\begin{table} [htb]
 \footnotesize
    \begin{tabular}{p{0.12\textwidth}p{0.13\textwidth}p{0.07\textwidth}p{0.32\textwidth}p{0.23\textwidth}}
    \hline
   Argument & Reasoning logic & Selected & Rationale & Typical evidence \\ 
    \hline
   Comparative & Inductive & \textbf{Yes} &
   Selected as the primary argument type for this case. Suitable when a validated human baseline exists, enabling empirical “no-worse-than” reasoning across fairness, consistency, and interpretability metrics. 
   & 
   Comparative evaluation results, expert review reports, red-team findings, and qualitative disagreement analysis. \\[6pt]

   Risk-based (Threshold-based) & Statistical & \textbf{Yes} &
   Selected as a complementary argument function to express quantitative confidence. Aggregates multiple proxy metrics and demonstrates non-inferiority ($\Delta \leq \delta$) with statistical confidence level ($\geq \alpha$). 
   & 
   Aggregated performance metrics, confidence intervals, bootstrapped significance tests, and uncertainty propagation analysis. \\[6pt]

   Normative & Deductive & No &
   Not selected as a primary argument in this case. The comparator (human-human baseline) and the associated procurement policies have already been formally accepted within governance processes; therefore, no additional normative justification is required. 
   & 
   (If applied) Regulatory standards, policy guidelines, and procurement framework documentation. \\ 
    \hline
    \end{tabular}
    \caption{Selected argument functions, reasoning logic, and rationale for the AI-based tender evaluation case.}
    \label{tab:argument_selection}
\end{table}

\textbf{Argument statements for the case}

With the selected argument functions, we formalised the following argument statements for the AI-based tender evaluation system. 

Comparative argument: 

\textit{"The AI–human tender evaluation process is at least as safe, fair, and reliable as the prior human–human process because its observed behaviour across multiple assurance dimensions (e.g., fairness, consistency, interpretability) is empirically comparable or superior under equivalent evaluation conditions. In particular, while the human–human review inconsistency averaged 3.0\%, the AI–human review inconsistency was observed at 2.8\%, yielding a marginal risk difference of $MR = \Delta R = R_{\text{AI+Human}} - R_{\text{Human+Human}} = -0.2\%$"}

This negative marginal difference indicates that the AI–human configuration performs at least as consistently as the human–human baseline, satisfying the comparative “no-worse-than” criterion for decision reliability.

Risk-based argument:

To support the comparative argument (e.g., decision consistency), the threshold-based argument is defined as follows. 

\textit{"The residual risk of decision inconsistency introduced by the AI reviewer remains below the predefined acceptance threshold $(\Delta \leq \delta)$ = 5\%."}

The threshold-based argument provides quantitative support for the comparative argument by statistically verifying that observed differences between AI–human and human–human evaluations remain within an acceptable tolerance range. It translates the qualitative claim of “no-worse-than” performance into measurable confidence bounds, thereby strengthening the empirical credibility of the primary argument.

\textbf{Evidence collection}


As mentioned earlier, there is no definitive ground truth in this case. Accordingly, the evaluation metrics in this case follow the principles outlined in Section \ref{sec:pattern} and rely on \textbf{predictability-based measures} such as consistency and consensus, rather than on absolute correctness. These metrics operationalise the \textbf{“no-ground-truth”} condition by quantifying relative coherence between human and AI reviewers.

To support the comparative and threshold-based arguments, evidence from \textit{Comparative benchmarking} was collected through controlled evaluation trials involving 200 synthetic tender cases.

The following statement clearly explains the assurance activities performed, summarises the tangible artefacts produced or collected, and explicitly links those artefacts to the argument they support.

\textit{"To substantiate the comparative and threshold-based arguments, a series of benchmarking and statistical evaluation activities were conducted. A synthetic dataset of 200 tender cases was prepared using historical scoring rubrics. Both the AI–human and human–human reviewer pairs independently assessed these cases following the same protocol. Inter-rater agreement, fairness deviation, and justification quality were measured.} 

\textit{The resulting data were analysed using bootstrapped significance tests and confidence-interval estimation to verify that observed differences $(\Delta = 2.8 \%, \delta = 5 \%, \alpha = 0.95)$ remained within acceptable tolerance levels."} 

The tangible evidence (corresponding artefacts) for this case includes: i) the comparative evaluation dataset, ii) the benchmarking and statistical analysis report, iii) expert review summaries, and iv) the evaluation protocol and computation scripts.

\textbf{Summary and conclusion}

This case applies the marginal-risk pattern without ground truth to an AI-assisted tender evaluation workflow. We advance a marginal safety claim supported by i) a Comparative argument (primary- inductive) and ii) a Threshold-based argument (supporting- statistical). 

Potential evidence for this case comes from the comparative benchmarking family, instantiated via predictability (decision stability/consistency) and capability (agreement, fairness, efficiency) metrics against the human–human baseline. These proxies triangulate safety in the absence of a definitive benchmark.

The results of this case study across 200 synthetic cases are interpreted as follows:

\begin{enumerate}
    \item \textbf{Observed marginal risk:}
    
    The differences (marginal risk: e.g., -0.2\% gap of inconsistency rate) clearly show that the new system (AI–human) is no worse (slightly better) than the existing system (human-human).

    The differences in decision inconsistency between the AI–human and human–human configurations ($\Delta R = -0.2\%$) show that the new system performs no worse and slightly better than the existing process.
    
    This indicates that the AI–human evaluation configuration achieves equivalent or improved performance (e.g., consistency and fairness).
    
    \item \textbf{Threshold validation:}
    
    In addition, the observed 2.8\% inconsistency rate of the AI–human system remained below the predefined acceptance threshold $(\delta = 5\%)$ with 95\% confidence.
    
    This statistically corroborates the comparative finding of “no-worse-than” performance and confirms that adding an AI reviewer does not increase residual decision risk beyond historical human variability.
\end{enumerate}

These results demonstrate that the AI–human evaluation process maintains decision reliability within the established human-human variability envelope.
By combining empirical comparison and statistical quantification, the AI safety case provides a defensible basis for concluding that the system operates within acceptable safety and fairness bounds.

The findings indicate that \textbf{the AI-based review process can be safely adopted} alongside or as a partial replacement for the existing human-human review workflow.
Adoption is expected to reduce reviewer workload and resource demand while maintaining consistency, fairness, and transparency in decision-making.
Continuous monitoring of agreement and fairness indicators is recommended to ensure that performance remains within the validated acceptance threshold.

\begin{figure} [htb]
    \centering
    \includegraphics[width=0.8\textwidth]{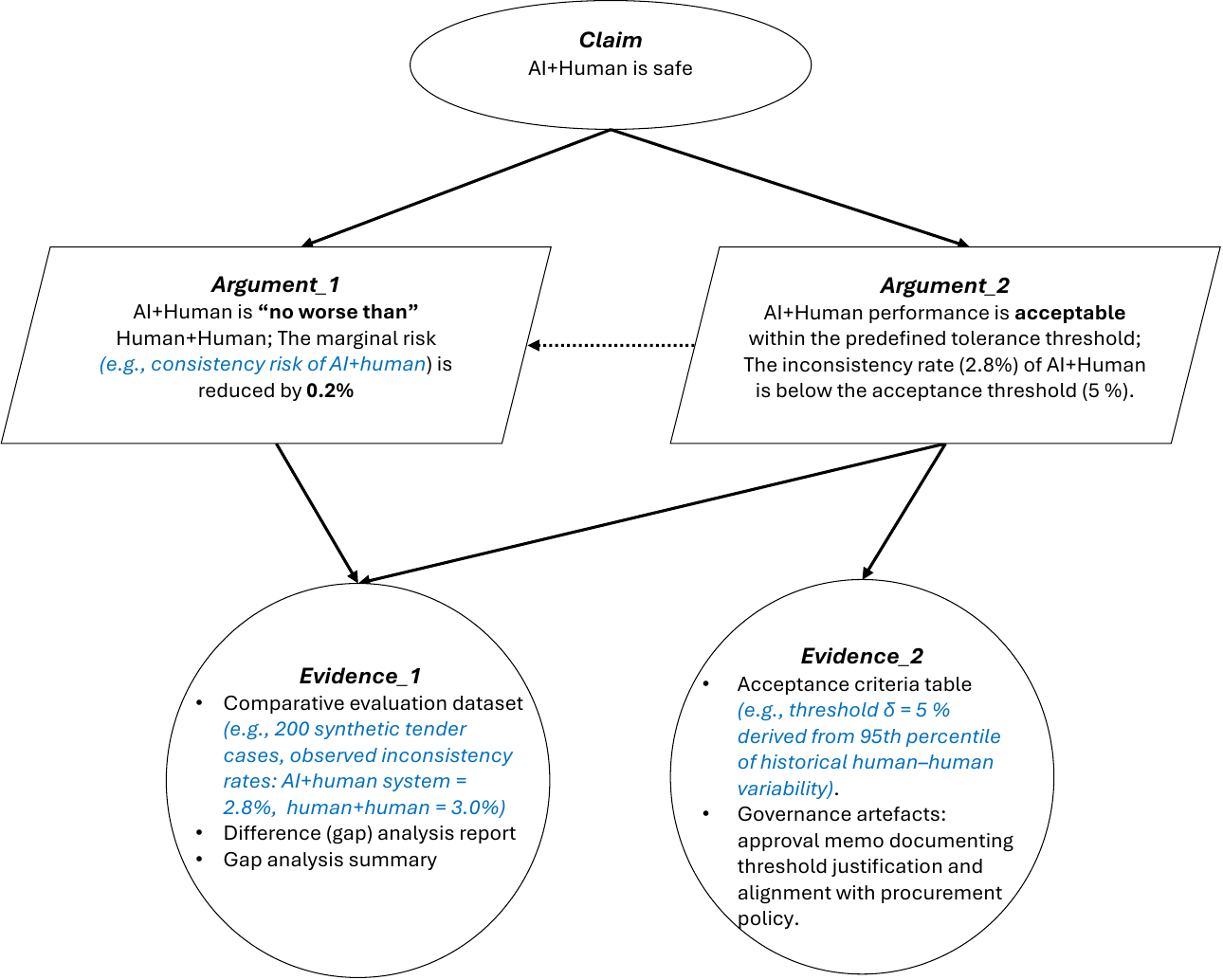}
    \caption{The AI safety case for AI-based tender evaluation system.}
    \label{fig:CAE for case study}
\end{figure}

Figure \ref{fig:CAE for case study} presents the AI safety case structure for this case study.

\section{Limitations and Future Work} \label{sec:limitation}

This study aims to establish reusable, composable templates for AI safety cases that reflect the empirical and evolving nature of AI systems. Several limitations remain.

\subsection{Scalability of dynamic safety cases}

Dynamic safety cases require continuous ingestion, verification, and version tracking of evidence. 
For continuously evolving systems, achieving through-life safety assurance is challenging due to the fundamental mismatch between rapid, non-deterministic AI evolution and the need for rigorous assurance documentation~\cite{Erfan2020dynamic,Shreyas2022dynamic,Laure2025dynamic}. 
Key limitations include continuous evidence overload from real-time operational data, the computational complexity of automatically updating argument structures when AI behaviour changes, and the lack of unified standards and robust automation infrastructure for managing dynamic evidence and argument revision. Implementing them at the scale of frontier AI systems will require automation infrastructure, tooling standards, and agreement on trusted data pipelines for live safety indicators.

\subsection{Interoperability across regulatory regimes}

Different jurisdictions emphasise distinct aspects of AI safety, such as capability thresholds, model access controls, or human oversight. 
Ensuring that reusable templates can be parameterised for diverse regulatory contexts without fragmentation is challenging due to divergent regulatory (e.g., EU AI Act~\cite{act2024eu}) philosophies, varied evidential standards and levels of assurance across sectors ~\cite{Leveson2011regulation}, and the difficulty of translating high-level legal requirements into structured, auditable technical claims within CAE templates~\cite{Natsuki2025regulation}. Achieving effective interoperability requires developing modular template components that can be selectively activated and customized for targeted jurisdictions and required assurance levels.

\subsection{Evaluation reproducibility}

Assurance for complex AI systems heavily depends on empirical evidence, such as red-teaming reports, adversarial testing results, and benchmark scores. However, the lack of standardization in these evaluation methods makes the resulting safety evidence difficult to verify or reproduce.
Additionally, fragmented evidence repositories prevent objective comparison of model safety performance against industry baselines, undermining the consistency of template-based safety claims. Improving reproducibility requires community consensus on standardized testing protocols, shared repositories of safety benchmarks and incident data, and open verification protocols to build trust in empirical assurance.

\subsection{Relationship to verification processes} 

This paper treats verification outcomes (e.g., test results, certificates, proofs, or monitoring summaries) as evidence inputs to an AI safety case, but does not model the verification processes that produce them. In particular, we do not characterise how behavioural properties are generated, checked, or enforced at design time or runtime. These concerns are increasingly addressed by verifiability-first AI engineering approaches \cite{zhuverifiability}, which focus on structuring systems to enable the verification of behavioural claims at scale. The contribution of this paper is complementary. It focuses on how verified artefacts, once available, are assembled, contextualised, and justified within a structured safety argument.

   


\subsection{Future directions}


Further research should operationalise the templates into a tool-supported Safety Case Pattern Library, enabling instantiation and continuous management across AI lifecycles. Key directions include formalising uncertainty propagation across reasoning modes, developing composite safety metrics for dynamic systems, and standardising comparator-based marginal-risk evaluations, all of which represent promising steps for both scientific and regulatory progress.
In addition, three priorities emerge for our next step work: First, using LLMs to automatically generate AI safety cases or CAE templates could significantly support dynamic safety case construction. Recent studies (e.g., S14,S51,S53,S74) show that LLMs can generate the structural components of safety case elements (e.g., CAEs) and accelerate early safety documentation, though they struggle to achieve high accuracy and provide traceable evidence without human review. This points to a hybrid human-AI assurance paradigm in which LLMs serve as drafting accelerators, while humans validate the results. Second, integrating AI risk management and assessment frameworks~\cite{NOPSEMA2025safetycase} and operational data, (e.g., guardrail logs~\cite{shamsujjoha_ICSA-2025}, evaluation results~\cite{balesni2024evaluations}, and AgentOps observability data~\cite{dong2024agentops2}) across AI safety case pipeline, could provide a more comprehensive evidence repository to support the full lifecycle of AI safety case construction, verification, and governance (as illustrated in Figure \ref{fig:pipeline} ecosystem pipeline). Third, developing systematic methods for mapping broad legal requirements into specific, measurable technical claims using CAE templates (e.g., ~\cite{Natsuki2025regulation}) would bridge the gap between regulatory mandates and technical assurance. These directions represent critical steps toward making rigorous, evidence-based safety assurance practical and scalable for frontier AI systems.

\section{Conclusion} 
\label{sec:concusion}
In this study, we presented a reusable safety-case template structure for AI systems. It uses a clear Claims-Arguments-Evidence (CAE) approach. It also provides AI-specific taxonomies for claim types, argument forms, and evidence families. These elements help practitioners build safety arguments in a consistent and reviewable way. Our proposed templates address recurring assurance issues in many AI deployments. We include end-to-end patterns for evaluation without ground truth, managing dynamic model updates, and making threshold-based risk decisions. This extends safety-case practice to areas where traditional safety cases often do not fit well. It also supports safety cases that can be checked and updated as the system changes. 

In addition, we embedded the templates into a continuous assurance pipeline. This pipeline links safety claims to live metrics and governance artifacts for ongoing monitoring. We applied the approach in a real-world case study on a government AI-based tender evaluation system. The case study shows how the templates can be used in practice and what evidence can be collected to support key claims.  Overall, this template-based approach provides a reusable foundation for AI safety cases that must evolve over time. However, it is not a complete solution. Key remaining challenges include scaling dynamic safety cases, supporting different regulatory settings, and improving the reproducibility of evaluations. 

\newpage
\section*{Appendix A ~ Seed Papers Sections}
\addcontentsline{toc}{section}{Appendix A ~  Seed Papers Sections}
\label{AppendixA}
 
\begin{table}[ht]
\centering
\small
\caption{Snowballing seed papers for AI safety case 
}
\label{seed-papers}
\begin{tabular}
{ p{0.3\textwidth} p{0.6\textwidth}}
\hline

 {Paper Title} &  {Key Focus \& Rationale Selection } \\
\hline

\textit{A Sketch of an AI Control Safety Case} &
Proposes a structured argument for controlling advanced AI systems, illustrating how control requirements map to claims, arguments, and evidence. \\

 \textit{GPT-4 and Safety Case Generation: An Exploratory Analysis} &
Explores whether GPT-4 can help generate safety case components, identifying strengths, limitations, and automation opportunities. \\

  \textit{Safety Case Template for Frontier AI: A Cyber Inability Argument} &
Introduces a template based on “cyber inability” reasoning, arguing that AI systems must be demonstrably unable to perform unsafe actions. \\

  \textit{Safety Case Templates for Autonomous Systems} &
Provides reusable safety case templates for autonomy, emphasising hazard analysis, assurance patterns, and argument structures. \\

  \textit{An Alignment Safety Case Sketch Based on Debate} &
Uses debate-based alignment methods as the core argument structure for an AI alignment safety case. \\

  \textit{The BIG Argument for AI Safety Cases} &
Presents the BIG (Balanced, Integrated and Grounded) argument pattern as a scalable foundation for justifying AI safety. \\

  \textit{Safety Cases: A Scalable Approach to Frontier AI Safety} &
Argues that safety cases can serve as a scalable governance tool for frontier AI models, outlining principles for systematic justification. \\
 
\textit{Safety Cases for Frontier AI} &
Discusses how safety cases can be adapted for frontier AI governance, emphasising transparency, auditability, and risk scenario coverage. \\

  \textit{Safety Cases: How to Justify the Safety of Advanced AI Systems} &
Introduces guidance for constructing safety cases for advanced AI, focusing on claims, evidence credibility, and structured reasoning. \\

\hline
\end{tabular}
\vspace{-1.3in}
\end{table}

\clearpage 
\begin{appendix}

\section*{Appendix B ~ List of Selected Studies}
\addcontentsline{toc}{section}{Appendix B ~ List of Selected Studies}

\label{AppendixB}

\renewcommand{\refname}{}


\renewcommand{\refname}{References}

\clearpage
\section*{Appendix C ~ Individual Quality Assessment Scores for Selected Studies}
\addcontentsline{toc}{section}{Appendix B ~ Individual quality assessment scores for selected studies}

\label{AppendixC}

\begin{table}[ht]
\centering
\label{QCassesSE1}
\begin{tabular}{|c|c|c|c|c|c|c|c|c|c|c|c|}
\hline
\textbf{Study No} &
  \textbf{QC1} &
  \textbf{QC2} &
  \textbf{QC3} &
  \textbf{QC4} &
  \textbf{QC5} &
  \textbf{QC6} &
  \textbf{QC7} &
  \textbf{QC8} &
  \textbf{QC9} &
  \textbf{QC10} &
  \textbf{QC11}  \\ \hline
  
\textbf{S1}	 	 &4	 &3	 &2	 &4	 &3	 &3	 &2	 &3	 &3	 &4	 &3	\\	\hline
\textbf{S2}	 	 &4	 &4	 &5	 &3	 &4	 &3	 &5	 &5	 &3	 &4	 &3	\\	\hline
\textbf{S3}	 	 &4	 &3	 &4	 &3	 &5	 &4	 &5	 &4	 &5	 &4	 &3	\\	\hline
\textbf{S4}	 	 &5	 &5	 &4	 &4	 &4	 &4	 &4	 &5	 &5	 &4	 &5	\\	\hline
\textbf{S5}	 	 &5	 &4	 &4	 &5	 &4	 &5	 &4	 &5	 &5	 &4	 &4	\\	\hline
\textbf{S6}	 	 &5	 &4	 &3	 &5	 &4	 &5	 &4	 &5	 &5	 &5	 &4	\\	\hline
\textbf{S7}	 	 &4	 &5	 &4	 &4	 &5	 &5	 &4	 &5	 &4	 &4	 &5	\\	\hline
\textbf{S8}	 	 &5	 &5	 &4	 &4	 &4	 &4	 &4	 &5	 &5	 &4	 &5	\\	\hline
\textbf{S9}	 	 &5	 &5	 &5	 &4	 &3	 &5	 &4	 &5	 &5	 &4	 &4	\\	\hline
\textbf{S10}	 &5	 &4	 &4	 &5	 &5	 &4	 &5	 &4	 &5	 &5	 &4	\\	\hline
\textbf{S11}	 &5	 &4	 &5	 &4	 &4	 &5	 &4	 &5	 &5	 &5	 &4	\\	\hline
\textbf{S12}	 &4	 &5	 &4	 &5	 &5	 &4	 &4	 &5	 &5	 &4	 &4	\\	\hline
\textbf{S13}	 &5	 &5	 &4	 &5	 &4	 &5	 &5	 &4	 &5	 &4	 &4	\\	\hline
\textbf{S14}	 &5	 &5	 &4	 &3	 &3	 &5	 &4	 &4	 &5	 &3	 &3	\\	\hline
\textbf{S15}	 &4	 &4	 &5	 &3	 &4	 &3	 &5	 &5	 &3	 &4	 &3	\\	\hline
\textbf{S16}	 &4	 &3	 &4	 &3	 &5	 &4	 &5	 &4	 &5	 &4	 &3	\\	\hline
\textbf{S17}	 &3	 &5	 &3	 &5	 &4	 &5	 &3	 &3	 &5	 &3	 &5	\\	\hline
\textbf{S18}	 &3	 &3	 &3	 &5	 &4	 &4	 &3	 &5	 &3	 &5	 &3	\\	\hline
\textbf{S19}	 &5	 &4	 &3	 &5	 &3	 &4	 &4	 &5	 &3	 &3	 &3	\\	\hline
\textbf{S20}	 &5	 &5	 &4	 &3	 &3	 &5	 &4	 &4	 &5	 &3	 &3	\\	\hline
\textbf{S21}	 &4	 &4	 &5	 &3	 &4	 &3	 &5	 &5	 &3	 &4	 &3	\\	\hline
\textbf{S22}	 &4	 &3	 &4	 &3	 &5	 &4	 &5	 &4	 &5	 &4	 &3	\\	\hline
\textbf{S23}	 &5	 &5	 &4	 &4	 &4	 &4	 &4	 &5	 &5	 &4	 &5	\\	\hline
\textbf{S24}	 &5	 &4	 &4	 &5	 &4	 &5	 &4	 &5	 &5	 &4	 &4	\\	\hline
\textbf{S25}	 &5	 &4	 &3	 &5	 &4	 &5	 &4	 &5	 &5	 &5	 &4	\\	\hline
\textbf{S26}	 &4	 &5	 &4	 &4	 &5	 &5	 &4	 &5	 &4	 &4	 &5	\\	\hline
\textbf{S27}	 &5	 &5	 &4	 &4	 &4	 &4	 &4	 &5	 &5	 &4	 &5	\\	\hline
\textbf{S28}	 &5	 &5	 &5	 &4	 &3	 &5	 &4	 &5	 &5	 &4	 &4	\\	\hline
\textbf{S29}	 &4	 &4	 &3	 &4	 &4	 &4	 &3	 &3	 &4	 &2	 &3	\\	\hline
\textbf{S30}	 &2	 &2	 &2	 &4	 &2	 &4	 &4	 &4	 &4	 &3	 &3	\\	\hline
\textbf{S31}	 &2	 &4	 &2	 &3	 &2	 &4	 &2	 &4	 &4	 &4	 &2	\\	\hline
\textbf{S32}	 &4	 &3	 &4	 &3	 &3	 &4	 &2	 &3	 &4	 &3	 &2	\\	\hline
\textbf{S33}	 &3	 &4	 &3	 &3	 &2	 &2	 &2	 &3	 &3	 &3	 &2	\\	\hline
\textbf{S34}	 &4	 &4	 &3	 &4	 &4	 &4	 &3	 &3	 &4	 &2	 &3	\\	\hline
\textbf{S35}	 &2	 &2	 &2	 &4	 &2	 &4	 &4	 &4	 &4	 &3	 &3	\\	\hline
\textbf{S36}	 &2	 &4	 &2	 &3	 &2	 &4	 &2	 &4	 &4	 &4	 &2	\\	\hline
\textbf{S37}	 &4	 &3	 &4	 &3	 &3	 &4	 &2	 &3	 &4	 &3	 &2	\\	\hline
\textbf{S38}	 &3	 &4	 &3	 &3	 &2	 &2	 &2	 &3	 &3	 &3	 &2	\\	\hline
\textbf{S39}	 &4	 &4	 &3	 &4	 &4	 &4	 &3	 &3	 &4	 &2	 &3	\\	\hline
\textbf{S40}	 &2	 &2	 &2	 &4	 &2	 &4	 &4	 &4	 &4	 &3	 &3	\\	\hline
\textbf{S41}	 &2	 &4	 &2	 &3	 &2	 &4	 &2	 &4	 &4	 &4	 &2	\\	\hline
\textbf{S42}	 &4	 &3	 &4	 &3	 &3	 &4	 &2	 &3	 &4	 &3	 &2	\\	\hline
\textbf{S43}	 &3	 &4	 &3	 &3	 &2	 &2	 &2	 &3	 &3	 &3	 &2	\\	\hline
\textbf{S44}	 &4	 &4	 &3	 &4	 &4	 &4	 &3	 &3	 &4	 &2	 &3	\\	\hline

\end{tabular}
\vspace{-2in}
\end{table}

\begin{table}[ht]
\centering
\label{QCassesSE2}
\begin{tabular}{|c|c|c|c|c|c|c|c|c|c|c|c|}
\hline
\textbf{Study No} &
  \textbf{QC1} &
  \textbf{QC2} &
  \textbf{QC3} &
  \textbf{QC4} &
  \textbf{QC5} &
  \textbf{QC6} &
  \textbf{QC7} &
  \textbf{QC8} &
  \textbf{QC9} &
  \textbf{QC10} &
  \textbf{QC11}  \\ \hline

\textbf{S45}	 &2	 &2	 &2	 &4	 &2	 &4	 &4	 &4	 &4	 &3	 &3	\\	\hline
\textbf{S46}	 &2	 &4	 &2	 &3	 &2	 &4	 &2	 &4	 &4	 &4	 &2	\\	\hline
\textbf{S47}	 &4	 &3	 &4	 &3	 &3	 &4	 &2	 &3	 &4	 &3	 &2	\\	\hline
\textbf{S48}	 &3	 &4	 &3	 &3	 &2	 &2	 &2	 &3	 &3	 &3	 &2	\\	\hline
\textbf{S49}	 &4	 &4	 &3	 &4	 &4	 &4	 &3	 &3	 &4	 &2	 &3	\\	\hline
\textbf{S50}	 &2	 &2	 &2	 &4	 &2	 &4	 &4	 &4	 &4	 &3	 &3	\\	\hline
\textbf{S51}	 &2	 &4	 &2	 &3	 &2	 &4	 &2	 &4	 &4	 &4	 &2	\\	\hline
\textbf{S52}	 &4	 &3	 &4	 &3	 &3	 &4	 &2	 &3	 &4	 &3	 &2	\\	\hline
\textbf{S53}	 &3	 &4	 &3	 &3	 &2	 &2	 &2	 &3	 &3	 &3	 &2	\\	\hline
\textbf{S54}	 &4	 &4	 &3	 &4	 &4	 &4	 &3	 &3	 &4	 &2	 &3	\\	\hline
\textbf{S55}	 &2	 &2	 &2	 &4	 &2	 &4	 &4	 &4	 &4	 &3	 &3	\\	\hline

\textbf{S56}	 &2	 &4	 &2	 &3	 &2	 &4	 &2	 &4	 &4	 &4	 &2	\\	\hline
  
\textbf{S57}	 &4	 &3	 &4	 &3	 &3	 &4	 &2	 &3	 &4	 &3	 &2	\\	\hline
\textbf{S58}	 &3	 &4	 &3	 &3	 &2	 &2	 &2	 &3	 &3	 &3	 &2	\\	\hline
\textbf{S59}	 &4	 &4	 &3	 &4	 &4	 &4	 &3	 &3	 &4	 &2	 &3	\\	\hline
\textbf{S60}	 &2	 &2	 &2	 &4	 &2	 &4	 &4	 &4	 &4	 &3	 &3	\\	\hline
\textbf{S61}	 &2	 &4	 &2	 &3	 &2	 &4	 &2	 &4	 &4	 &4	 &2	\\	\hline

\textbf{S62}	 &4	 &3	 &4	 &3	 &3	 &4	 &2	 &3	 &4	 &3	 &2	\\	\hline
\textbf{S63}	 &3	 &4	 &3	 &3	 &2	 &2	 &2	 &3	 &3	 &3	 &2	\\	\hline
\textbf{S64}	 &4	 &4	 &3	 &4	 &4	 &4	 &3	 &3	 &4	 &2	 &3	\\	\hline
\textbf{S65}	 &2	 &2	 &2	 &4	 &2	 &4	 &4	 &4	 &4	 &3	 &3	\\	\hline
\textbf{S66}	 &2	 &4	 &2	 &3	 &2	 &4	 &2	 &4	 &4	 &4	 &2	\\	\hline
\textbf{S67}	 &4	 &3	 &4	 &3	 &3	 &4	 &2	 &3	 &4	 &3	 &2	\\	\hline
\textbf{S68}	 &3	 &4	 &3	 &3	 &2	 &2	 &2	 &3	 &3	 &3	 &2	\\	\hline
\textbf{S69}	 &4	 &4	 &3	 &4	 &4	 &4	 &3	 &3	 &4	 &2	 &3	\\	\hline
\textbf{S70}	 &2	 &2	 &2	 &4	 &2	 &4	 &4	 &4	 &4	 &3	 &3	\\	\hline
\textbf{S71}	 &2	 &4	 &2	 &3	 &2	 &4	 &2	 &4	 &4	 &4	 &2	\\	\hline
\textbf{S72}	 &4	 &3	 &4	 &3	 &3	 &4	 &2	 &3	 &4	 &3	 &2	\\	\hline
\textbf{S73}	 &3	 &4	 &3	 &3	 &2	 &2	 &2	 &3	 &3	 &3	 &2	\\	\hline
\textbf{S74}	 &4	 &4	 &3	 &4	 &4	 &4	 &3	 &3	 &4	 &2	 &3	\\	\hline
\textbf{S75}	 &2	 &2	 &2	 &4	 &2	 &4	 &4	 &4	 &4	 &3	 &3	\\	\hline
\textbf{S76}	 &2	 &4	 &2	 &3	 &2	 &4	 &2	 &4	 &4	 &4	 &2	\\	\hline
\textbf{S77}	 &4	 &3	 &4	 &3	 &3	 &4	 &2	 &3	 &4	 &3	 &2	\\	\hline
\textbf{S78}	 &3	 &4	 &3	 &3	 &2	 &2	 &2	 &3	 &3	 &3	 &2	\\	\hline
\textbf{S79}	 &4	 &4	 &3	 &4	 &4	 &4	 &3	 &3	 &4	 &2	 &3	\\	\hline
\textbf{S80}	 &2	 &2	 &2	 &4	 &2	 &4	 &4	 &4	 &4	 &3	 &3	\\	\hline
\textbf{S81}	 &2	 &4	 &2	 &3	 &2	 &4	 &2	 &4	 &4	 &4	 &2	\\	\hline
\textbf{S82}	 &4	 &3	 &4	 &3	 &3	 &4	 &2	 &3	 &4	 &3	 &2	\\	\hline
\textbf{S83}	 &3	 &4	 &3	 &3	 &2	 &2	 &2	 &3	 &3	 &3	 &2	\\	\hline
\textbf{S84}	 &4	 &4	 &3	 &4	 &4	 &4	 &3	 &3	 &4	 &2	 &3	\\	\hline
\textbf{S85}	 &2	 &2	 &2	 &4	 &2	 &4	 &4	 &4	 &4	 &3	 &3	\\	\hline
\textbf{S86}	 &2	 &4	 &2	 &3	 &2	 &4	 &2	 &4	 &4	 &4	 &2	\\	\hline
\textbf{S87}	 &4	 &3	 &4	 &3	 &3	 &4	 &2	 &3	 &4	 &3	 &2	\\	\hline
\textbf{S88}	 &3	 &4	 &3	 &3	 &2	 &2	 &2	 &3	 &3	 &3	 &2	\\	\hline
\textbf{S89}	 &4	 &4	 &3	 &4	 &4	 &4	 &3	 &3	 &4	 &2	 &3	\\	\hline
\textbf{S90}	 &2	 &2	 &2	 &4	 &2	 &4	 &4	 &4	 &4	 &3	 &3	\\	\hline

\end{tabular}
\vspace{-0.15in}
\end{table}

\begin{table}[ht]
\centering
\label{QCassesSE3}
\begin{tabular}{|c|c|c|c|c|c|c|c|c|c|c|c|}
\hline
\textbf{Study No} &
  \textbf{QC1} &
  \textbf{QC2} &
  \textbf{QC3} &
  \textbf{QC4} &
  \textbf{QC5} &
  \textbf{QC6} &
  \textbf{QC7} &
  \textbf{QC8} &
  \textbf{QC9} &
  \textbf{QC10} &
  \textbf{QC11}  \\ \hline

\textbf{S91}	 &2	 &4	 &2	 &3	 &2	 &4	 &2	 &4	 &4	 &4	 &2	\\	\hline

\textbf{S92}	 &4	 &3	 &4	 &3	 &3	 &4	 &2	 &3	 &4	 &3	 &2	\\	\hline
\textbf{S93}	 &3	 &4	 &3	 &3	 &2	 &2	 &2	 &3	 &3	 &3	 &2	\\	\hline
\textbf{S94}	 &4	 &4	 &3	 &4	 &4	 &4	 &3	 &3	 &4	 &2	 &3	\\	\hline
\textbf{S95}	 &2	 &2	 &2	 &4	 &2	 &4	 &4	 &4	 &4	 &3	 &3	\\	\hline
\textbf{S96}	 &2	 &4	 &2	 &3	 &2	 &4	 &2	 &4	 &4	 &4	 &2	\\	\hline
\textbf{S97}	 &4	 &3	 &4	 &3	 &3	 &4	 &2	 &3	 &4	 &3	 &2	\\	\hline
\textbf{S98}	 &3	 &4	 &3	 &3	 &2	 &2	 &2	 &3	 &3	 &3	 &2	\\	\hline
\textbf{S99}	 &4	 &4	 &3	 &4	 &4	 &4	 &3	 &3	 &4	 &2	 &3	\\	\hline
\textbf{S100}	 &2	 &2	 &2	 &4	 &2	 &4	 &4	 &4	 &4	 &3	 &3	\\	\hline
\textbf{S101}	 &2	 &4	 &2	 &3	 &2	 &4	 &2	 &4	 &4	 &4	 &2	\\	\hline
\textbf{S102}	 &4	 &3	 &4	 &3	 &3	 &4	 &2	 &3	 &4	 &3	
&2	\\	\hline

\textbf{S103}	 &3	 &4	 &3	 &3	 &2	 &2	 &2	 &3	 &3	 &3	 &2	\\	\hline
\textbf{S104}	 &4	 &4	 &3	 &4	 &4	 &4	 &3	 &3	 &4	 &2	 &3	\\	\hline
\textbf{S105}	 &2	 &2	 &2	 &4	 &2	 &4	 &4	 &4	 &4	 &3	 &3	\\	\hline

\textbf{S106}	 &2	 &4	 &2	 &3	 &2	 &4	 &2	 &4	 &4	 &4	 &2	\\	\hline
\textbf{S107}	 &4	 &3	 &4	 &3	 &3	 &4	 &2	 &3	 &4	 &3	 &2	\\	\hline
\textbf{S108}	 &4	 &4	 &3	 &4	 &4	 &4	 &3	 &3	 &4	 &2	 &3	\\	\hline
\textbf{S109}	 &2	 &2	 &2	 &4	 &2	 &4	 &4	 &4	 &4	 &3	 &3	\\	\hline
\textbf{S110}	 &2	 &4	 &2	 &3	 &2	 &4	 &2	 &4	 &4	 &4	 &2	\\	\hline
\textbf{S111}	 &4	 &3	 &4	 &3	 &3	 &4	 &2	 &3	 &4	 &3	 &2	\\	\hline
\textbf{S112}	 &3	 &4	 &3	 &3	 &2	 &2	 &2	 &3	 &3	 &3	 &2	\\	\hline

\end{tabular}
\vspace{-0.15in}
\end{table}

\end{appendix}

\clearpage

\bibliographystyle{IEEEtran} 
\bibliography{Reference}

\begin{thebibliography}{112}{
\expandafter\ifx\csname url\endcsname\relax
  \def\url#1{\texttt{#1}}\fi
\expandafter\ifx\csname urlprefix\endcsname\relax\def\urlprefix{URL }\fi
\expandafter\ifx\csname href\endcsname\relax
  \def\href#1#2{#2} \def\path#1{#1}\fi


\bibitem[S1]{S2-causal-model}
S.~Burton, ``A causal model of safety assurance for machine learning,'' \emph{arXiv preprint arXiv:2201.05451}, 2022.

\bibitem[S2]{S3-co-simulation}
D.~Tola and P.~G. Larsen, ``A co-simulation based approach for developing safety-critical systems,'' in \emph{Proceedings of the 18th International Overture Workshop}, vol.~65, 2021.

\bibitem[S3]{S4-consistent-ai-argument}
A.~Rudolph, S.~Voget, and J.~Mottok, ``A consistent safety case argumentation for artificial intelligence in safety related automotive systems,'' in \emph{ERTS 2018}, 2018.

\bibitem[S4]{S5-deductive-safety-contract}
B.~Herd, J.-V. Zacchi, and S.~Burton, ``A deductive approach to safety assurance: Formalising safety contracts with subjective logic,'' in \emph{International Conference on Computer Safety, Reliability, and Security}.\hskip 1em plus 0.5em minus 0.4em\relax Springer Nature Switzerland Cham, 2024, pp. 213--226.

\bibitem[S5]{S6-deployment-ai-diagnosis}
Y.~Jia, C.~Verrill, K.~White, M.~Dolton, M.~Horton, M.~Jafferji, and I.~Habli, ``A deployment safety case for ai-assisted prostate cancer diagnosis,'' \emph{Computers in biology and medicine}, vol. 192, p. 110237, 2025.

\bibitem[S6]{S7-formal-basis-pattern}
E.~Denney and G.~Pai, ``A formal basis for safety case patterns,'' in \emph{International Conference on Computer Safety, Reliability, and Security}.\hskip 1em plus 0.5em minus 0.4em\relax Springer, 2013, pp. 21--32.

\bibitem[S7]{S8-formal-methods-evidence}
J.~Krook, Y.~Selvaraj, W.~Ahrendt, and M.~Fabian, ``A formal-methods approach to provide evidence in automated-driving safety cases,'' \emph{arXiv preprint arXiv:2210.07798}, 2022.

\bibitem[S8]{S9-gsn-ai-regulation}
N.~Hayama, Y.~Yamagata, H.~Nishihara, and Y.~Matsuno, ``A gsn-based requirement analysis of the eu {AI} regulation,'' in \emph{International Conference on Computer Safety, Reliability, and Security}.\hskip 1em plus 0.5em minus 0.4em\relax Springer Nature Switzerland Cham, 2025, pp. 183--196.

\bibitem[S9]{S10-safety-case-methodology}
P.~Bishop and R.~Bloomfield, ``A methodology for safety case development,'' in \emph{Safety and Reliability}, vol.~20, no.~1.\hskip 1em plus 0.5em minus 0.4em\relax Taylor \& Francis, 2000, pp. 34--42.

\bibitem[S10]{S11-ethics-assurance-pattern}
Z.~Porter, I.~Habli, J.~McDermid, and M.~Kaas, ``A principles-based ethics assurance argument pattern for {AI} and autonomous systems,'' \emph{AI and Ethics}, vol.~4, no.~2, pp. 593--616, 2024.

\bibitem[S11]{S12-prisma-bibliometric}
O.~Odu, A.~B. Belle, S.~Wang, and K.~K. Shahandashti, ``A prisma-driven bibliometric analysis of the scientific literature on assurance case patterns,'' \emph{arXiv preprint arXiv:2407.04961}, 2024.

\bibitem[S12]{S13-prisma-weakeners}
K.~K. Shahandashti, A.~B. Belle, T.~C. Lethbridge, O.~Odu, and M.~Sivakumar, ``A prisma-driven systematic mapping study on system assurance weakeners,'' \emph{Information and Software Technology}, vol. 175, p. 107526, 2024.

\bibitem[S13]{S14-safety-argument-fragment}
M.~Gyllenhammar, G.~R.~d. Campos, and M.~T{\"o}rngren, ``A safety argument fragment towards safe deployment of performant automated driving systems,'' in \emph{International Conference on Computer Safety, Reliability, and Security}.\hskip 1em plus 0.5em minus 0.4em\relax Springer, 2025, pp. 197--210.

\bibitem[S14]{S15-llm-corrigibility}
R.~Potham, ``A safety case for a deployed llm: Corrigibility as a singular target via debate,'' 2025.

\bibitem[S15]{S16-ml-pattern}
E.~Wozniak, C.~C{\^a}rlan, E.~Acar-Celik, and H.~J. Putzer, ``A safety case pattern for systems with machine learning components,'' in \emph{International Conference on Computer Safety, Reliability, and Security}.\hskip 1em plus 0.5em minus 0.4em\relax Springer, 2020, pp. 370--382.

\bibitem[S16]{S18-ai-control-sketch}
T.~Korbak, J.~Clymer, B.~Hilton, B.~Shlegeris, and G.~Irving, ``A sketch of an {AI} control safety case,'' \emph{arXiv preprint arXiv:2501.17315}, 2025.

\bibitem[S17]{S19-access-engineering}
R.~Wei, S.~Foster, H.~Mei, F.~Yan, R.~Yang, I.~Habli, C.~O’Halloran, N.~Tudor, T.~Kelly, and Y.~Nemouchi, ``Access: Assurance case centric engineering of safety--critical systems,'' \emph{Journal of Systems and Software}, vol. 213, p. 112034, 2024.

\bibitem[S18]{S20-uncertainty-assurance}
S.~Burton and B.~Herd, ``Addressing uncertainty in the safety assurance of machine-learning,'' \emph{Frontiers in computer science}, vol.~5, p. 1132580, 2023.

\bibitem[S19]{S22-agile-aml}
V.~J. Hodge and M.~Osborne, ``Agile development for safety assurance of machine learning in autonomous systems (agileamlas),'' \emph{Array}, p. 100482, 2025.

\bibitem[S20]{S23-agile-autonomous}
B.~Kaiser, M.~Soden, R.~Diefenbach, and E.~Holz, ``An agile approach to safety cases for autonomous systems through model-based engineering and simulation,'' in \emph{Proc. 33rd Saf. Crit. Syst. Symp}, 2025.

\bibitem[S21]{S24-alignment-debate}
M.~D. Buhl, J.~Pfau, B.~Hilton, and G.~Irving, ``An alignment safety case sketch based on debate,'' \emph{arXiv preprint arXiv:2505.03989}, 2025.

\bibitem[S22]{S25-interpretability-pattern}
F.~R. Ward and I.~Habli, ``An assurance case pattern for the interpretability of machine learning in safety-critical systems,'' in \emph{International Conference on Computer Safety, Reliability, and Security}.\hskip 1em plus 0.5em minus 0.4em\relax Springer, 2020, pp. 395--407.

\bibitem[S23]{S26-misuse-safeguard}
J.~Clymer, J.~Weinbaum, R.~Kirk, K.~Mai, S.~Zhang, and X.~Davies, ``An example safety case for safeguards against misuse,'' \emph{arXiv preprint arXiv:2505.18003}, 2025.

\bibitem[S24]{S27-evidence-certification}
S.~Nair, J.~L. De~La~Vara, M.~Sabetzadeh, and L.~Briand, ``An extended systematic literature review on provision of evidence for safety certification,'' \emph{Information and Software Technology}, vol.~56, no.~7, pp. 689--717, 2014.

\bibitem[S25]{S28-ul4600-applicability}
U.~D. Ferrell and A.~H.~A. Anderegg, ``Applicability of ul 4600 to unmanned aircraft systems (uas) and urban air mobility (uam),'' in \emph{2020 AIAA/IEEE 39th Digital Avionics Systems Conference (DASC)}.\hskip 1em plus 0.5em minus 0.4em\relax IEEE, 2020, pp. 1--7.

\bibitem[S26]{S29-concept-model}
J.~P.~C. de~Araujo, B.~V. Balu, E.~Reichmann, J.~Kelly, S.~Kugele, N.~Mata, and L.~Grunske, ``Applying concept-based models for enhanced safety argumentation,'' in \emph{2024 IEEE 35th International Symposium on Software Reliability Engineering (ISSRE)}.\hskip 1em plus 0.5em minus 0.4em\relax IEEE, 2024, pp. 272--283.

\bibitem[S27]{S30-arguing-safety}
T.~P. Kelly, ``Arguing safety-a systematic approach to safety case management,'' \emph{Department of Computer Science, The University of York}, 1998.

\bibitem[S28]{S31-confidence-frontier}
S.~Barrett, P.~Fox, J.~Krook, T.~Mondal, S.~Mylius, and A.~Tlaie, ``Assessing confidence in frontier {AI} safety cases,'' \emph{arXiv preprint arXiv:2502.05791}, 2025.

\bibitem[S29]{S33-cps-dnn-pattern}
R.~Kaur, R.~Ivanov, M.~Cleaveland, O.~Sokolsky, and I.~Lee, ``Assurance case patterns for cyber-physical systems with deep neural networks,'' in \emph{International Conference on Computer Safety, Reliability, and Security}.\hskip 1em plus 0.5em minus 0.4em\relax Springer, 2020, pp. 82--97.

\bibitem[S30]{S34-assurance-template-principles}
T.~Chowdhury, ``Assurance case templates: Principles for their development and criteria for their evaluation,'' Ph.D. dissertation, McMaster University, 2021.

\bibitem[S31]{S35-ml-verification}
V.~Mussot, E.~Jenn, F.~Chenevier, R.~C. Laguna, Y.~I. Messaoud, J.-L. Farges, A.~F. Pires, F.~Latombe, and S.~Creff, ``Assurance cases to face the complexity of ml-based systems verification,'' in \emph{Embedded Real Time System Congress, ERTS'24}, 2024.

\bibitem[S32]{S36-ai-dependability}
R.~Bloomfield and J.~Rushby, ``Assurance of {AI} systems from a dependability perspective,'' \emph{arXiv preprint arXiv:2407.13948}, 2024.

\bibitem[S33]{S37-aviation-ml-design}
E.~Denney and G.~Pai, ``Assurance-driven design of machine learning-based functionality in an aviation systems context,'' in \emph{2023 IEEE/AIAA 42nd Digital Avionics Systems Conference (DASC)}.\hskip 1em plus 0.5em minus 0.4em\relax IEEE, 2023, pp. 1--10.

\bibitem[S34]{S40-modular-assurance}
A.~Wardzi{\'n}ski and A.~Jarz{\k{e}}bowicz, ``Automated generation of modular assurance cases with the system assurance reference model,'' \emph{Formal Aspects of Computing}, vol.~36, no.~4, pp. 1--29, 2024.

\bibitem[S35]{S41-llm-instantiate-pattern}
O.~Odu, A.~B. Belle, S.~Wang, S.~Kpodjedo, T.~C. Lethbridge, and H.~Hemmati, ``Automatic instantiation of assurance cases from patterns using large language models,'' \emph{Journal of Systems and Software}, vol. 222, p. 112353, 2025.

\bibitem[S36]{S42-ml-change-impact}
C.~C{\^a}rlan, L.~Gauerhof, B.~Gallina, and S.~Burton, ``Automating safety argument change impact analysis for machine learning components,'' in \emph{2022 IEEE 27th Pacific Rim International Symposium on Dependable Computing (PRDC)}.\hskip 1em plus 0.5em minus 0.4em\relax IEEE, 2022, pp. 43--53.

\bibitem[S37]{S43-llm-assurance-balance}
S.~Diemert, E.~Cyffka, N.~Anwari, O.~Foster, T.~Viger, L.~Millet, and J.~Joyce, ``Balancing the risks and benefits of using large language models to support assurance case development,'' in \emph{International Conference on Computer Safety, Reliability, and Security}.\hskip 1em plus 0.5em minus 0.4em\relax Springer, 2025, pp. 209--225.

\bibitem[S38]{S44-checkable-safety-arguments}
C.~C{\^a}rlan, ``Checkable safety arguments-a modeling framework supporting the maintenance of safety arguments consistent with system development artifacts,'' Ph.D. dissertation, Technische Universit{\"a}t M{\"u}nchen, 2025.

\bibitem[S39]{S45-confidence-structure}
Y.~Idmessaoud, D.~Dubois, and J.~Guiochet, ``Confidence assessment in safety argument structure-quantitative vs. qualitative approaches,'' \emph{International Journal of Approximate Reasoning}, vol. 165, p. 109100, 2024.

\bibitem[S40]{S46-ml-autonomous-driving}
M.~Sivakumar, ``Design and automatic generation of safety cases of ml-enabled autonomous driving systems,'' Master's thesis, York University, 2024.

\bibitem[S41]{S47-rl-autonomous-vehicle}
M.~Sivakumar, A.~B. Belle, J.~Shan, O.~Odu, and M.~Yuan, ``Design of the safety case of the reinforcement learning-enabled component of a quanser autonomous vehicle,'' in \emph{2024 IEEE 32nd International Requirements Engineering Conference Workshops (REW)}.\hskip 1em plus 0.5em minus 0.4em\relax IEEE, 2024, pp. 57--67.

\bibitem[S42]{S48-dynamic-modular-runtime}
E.~Mirzaei, C.~CARLAN, C.~Thomas, and B.~Gallina, ``Design-time specification of dynamic modular safety cases in support of run-time safety assessment,'' in \emph{Proceedings of the 30th Safety-Critical Systems Symposium (SCSC)}, 2022.

\bibitem[S43]{S49-compelling-safety-cases}
R.~Hawkins, ``Developing compelling safety cases,'' \emph{arXiv preprint arXiv:2502.00911}, 2025.

\bibitem[S44]{S50-dynamic-assurance}
E.~Asaadi, E.~Denney, J.~Menzies, G.~J. Pai, and D.~Petroff, ``Dynamic assurance cases: a pathway to trusted autonomy,'' \emph{Computer}, vol.~53, no.~12, pp. 35--46, 2020.

\bibitem[S45]{S51-dynamic-safety-cps-}
S.~Ramakrishna, ``Dynamic safety assurance of autonomous cyber-physical systems,'' Ph.D. dissertation, Vanderbilt University, 2022.

\bibitem[S46]{S52-frontier-ai-dynamic-}
C.~C{\^a}rlan, F.~Gomez, Y.~Mathew, K.~Krishna, R.~King, P.~Gebauer, and B.~R. Smith, ``Dynamic safety cases for frontier ai,'' \emph{arXiv preprint arXiv:2412.17618}, 2024.

\bibitem[S47]{S53-smirk-safe}
M.~Borg, J.~Henriksson, K.~Socha, O.~Lennartsson, E.~Sonnsj{\"o}~L{\"o}negren, T.~Bui, P.~Tomaszewski, S.~R. Sathyamoorthy, S.~Brink, and M.~Helali~Moghadam, ``Ergo, smirk is safe: a safety case for a machine learning component in a pedestrian automatic emergency brake system,'' \emph{Software quality journal}, vol.~31, no.~2, pp. 335--403, 2023.

\bibitem[S48]{S54-ethical-assurance}
C.~Burr and D.~Leslie, ``Ethical assurance: a practical approach to the responsible design, development, and deployment of data-driven technologies,'' \emph{AI and Ethics}, vol.~3, no.~1, pp. 73--98, 2023.

\bibitem[S49]{S55-autonomous-railway}
J.~Ro{\ss}bach, O.~De~Candido, A.~Hammam, and M.~Leuschel, ``Evaluating ai-based components in autonomous railway systems: A methodology,'' in \emph{German Conference on Artificial Intelligence (K{\"u}nstliche Intelligenz)}.\hskip 1em plus 0.5em minus 0.4em\relax Springer, 2024, pp. 190--203.

\bibitem[S50]{S57-explainable-compliance}
F.~Ikhwantri and D.~Marijan, ``Explainable compliance detection with multi-hop natural language inference on assurance case structure,'' \emph{arXiv preprint arXiv:2506.08713}, 2025.

\bibitem[S51]{S58-gpt4-generation}
M.~Sivakumar, A.~B. Belle, J.~Shan, and K.~K. Shahandashti, ``Exploring the capabilities of large language models for the generation of safety cases: the case of gpt-4,'' in \emph{2024 IEEE 32nd International Requirements Engineering Conference Workshops (REW)}.\hskip 1em plus 0.5em minus 0.4em\relax IEEE, 2024, pp. 35--45.

\bibitem[S52]{S60-safety-contract-fragment}
I.~Sljivo, B.~Gallina, J.~Carlson, and H.~Hansson, ``Generation of safety case argument-fragments from safety contracts,'' in \emph{International Conference on Computer Safety, Reliability, and Security}.\hskip 1em plus 0.5em minus 0.4em\relax Springer, 2014, pp. 170--185.

\bibitem[S53]{S61-gpt4-safety-case}
M.~Sivakumar, A.~B. Belle, J.~Shan, and K.~K. Shahandashti, ``Gpt-4 and safety case generation: An exploratory analysis,'' \emph{arXiv preprint arXiv:2312.05696}, 2023.

\bibitem[S54]{S62-gsn-autonomous-trains}
M.~Chelouati, A.~Boussif, J.~Beugin, and E.-M. El~Koursi, ``Graphical safety assurance case using goal structuring notation (gsn)—challenges, opportunities and a framework for autonomous trains,'' \emph{Reliability Engineering \& System Safety}, vol. 230, p. 108933, 2023.

\bibitem[S55]{S63-formal-verification-integration}
I.~Sljivo, E.~Denney, and J.~Menzies, ``Guided integration of formal verification in assurance cases,'' in \emph{International Conference on Formal Engineering Methods}.\hskip 1em plus 0.5em minus 0.4em\relax Springer, 2023, pp. 172--190.

\bibitem[S56]{S64-runtime-monitoring}
R.~Hawkins and P.~Ryan~Conmy, ``Identifying run-time monitoring requirements for autonomous systems through the analysis of safety arguments,'' in \emph{International Conference on Computer Safety, Reliability, and Security}.\hskip 1em plus 0.5em minus 0.4em\relax Springer, 2023, pp. 11--24.

\bibitem[S57]{S65-health-it-safety-cases}
G.~Despotou, S.~White, T.~Kelly, and M.~Ryan, ``Introducing safety cases for health it,'' in \emph{2012 4th International Workshop on Software Engineering in Health Care (SEHC)}.\hskip 1em plus 0.5em minus 0.4em\relax IEEE, 2012, pp. 44--50.

\bibitem[S58]{S66-software-standards}
M.~J. Squair, ``Issues in the application of software safety standards,'' in \emph{ACM International Conference Proceeding Series}, vol. 162, 2006, pp. 13--26.

\bibitem[S59]{S67-fairness-assurance}
A.~Sabuncuoglu, C.~Burr, and C.~Maple, ``Justified evidence collection for argument-based {AI} fairness assurance,'' in \emph{Proceedings of the 2025 ACM Conference on Fairness, Accountability, and Transparency}, 2025, pp. 18--28.

\bibitem[S60]{S68-artifact-trees}
A.~Agrawal, S.~Khoshmanesh, M.~Vierhauser, M.~Rahimi, J.~Cleland-Huang, and R.~Lutz, ``Leveraging artifact trees to evolve and reuse safety cases,'' in \emph{2019 IEEE/ACM 41st International Conference on Software Engineering (ICSE)}.\hskip 1em plus 0.5em minus 0.4em\relax IEEE, 2019, pp. 1222--1233.

\bibitem[S61]{S70-llm-mitigation-generation}
Y.~Fujiwara, T.~Tuchida, R.~Miyata, H.~Washizaki, and N.~Ubayashi, ``Llm-based automated mitigation and assurance case generation against threats to {AI} systems,'' in \emph{2025 IEEE Conference on Artificial Intelligence (CAI)}.\hskip 1em plus 0.5em minus 0.4em\relax IEEE, 2025, pp. 906--911.

\bibitem[S62]{S71-apollo-safety-case}
O.~Odu, A.~B. Belle, and S.~Wang, ``Llm-based safety case generation for baidu apollo: Are we there yet?'' in \emph{2025 IEEE/ACM 4th International Conference on {AI} Engineering--Software Engineering for {AI} (CAIN)}.\hskip 1em plus 0.5em minus 0.4em\relax IEEE, 2025, pp. 222--233.

\bibitem[S63]{S72-highly-automated-driving}
S.~Burton, L.~Gauerhof, and C.~Heinzemann, ``Making the case for safety of machine learning in highly automated driving,'' in \emph{International Conference on Computer Safety, Reliability, and Security}.\hskip 1em plus 0.5em minus 0.4em\relax Springer, 2017, pp. 5--16.

\bibitem[S64]{S73-misaligned-ai-risk}
R.~Dassanayake, M.~Demetroudi, J.~Walpole, L.~Lentati, J.~R. Brown, and E.~J. Young, ``Manipulation attacks by misaligned ai: Risk analysis and safety case framework,'' \emph{arXiv preprint arXiv:2507.12872}, 2025.

\bibitem[S65]{S74-confidence-measure}
C.-L. Lin, W.~Shen, S.~Drager, and B.~Cheng, ``Measure confidence of assurance cases in safety-critical domains,'' in \emph{Proceedings of the 40th International Conference on Software Engineering: New Ideas and Emerging Results}, 2018, pp. 13--16.

\bibitem[S66]{S75-psm-safety}
J.~Murdoch, G.~Clark, A.~Powell, and P.~Caseley, ``Measuring safety: applying psm to the system safety domain,'' in \emph{Proceedings of the 8th Australian workshop on Safety critical systems and software-Volume 33}, 2003, pp. 47--55.

\bibitem[S67]{S76-modalas-uncertainty}
M.~A. Langford, K.~H. Chan, J.~E. Fleck, P.~K. McKinley, and B.~H. Cheng, ``MoDALAS: addressing assurance for learning-enabled autonomous systems in the face of uncertainty,'' \emph{Software and systems modeling}, vol.~22, no.~5, pp. 1543--1563, 2023.

\bibitem[S68]{S77-security-assurance}
L.~Nascimento, A.~L. de~Oliveira, R.~Villela, E.~F. Silva, R.~Wei, R.~Hawkins, and T.~Kelly, ``Model-based security assurance cases for open and adaptive cyber-physical systems,'' in \emph{International Conference on Advanced Information Networking and Applications}.\hskip 1em plus 0.5em minus 0.4em\relax Springer, 2025, pp. 326--340.

\bibitem[S69]{S78-model-connected-case}
A.~Retouniotis, Y.~Papadopoulos, I.~Sorokos, D.~Parker, N.~Matragkas, and S.~Sharvia, ``Model-connected safety cases,'' in \emph{International Symposium on Model-Based Safety and Assessment}.\hskip 1em plus 0.5em minus 0.4em\relax Springer, 2017, pp. 50--63.

\bibitem[S70]{S79-goal-based-management}
D.~H. Becht, ``Moving towards goal-based safety management,'' in \emph{Proceedings of the Australian System Safety Conference-Volume 133}, 2011, pp. 19--26.

\bibitem[S71]{S80-nn-dependability}
C.-H. Cheng, C.-H. Huang, and G.~N{\"u}hrenberg, ``nn-dependability-kit: Engineering neural networks for safety-critical autonomous driving systems,'' in \emph{2019 IEEE/ACM International Conference on Computer-Aided Design (ICCAD)}.\hskip 1em plus 0.5em minus 0.4em\relax IEEE, 2019, pp. 1--6.

\bibitem[S72]{S81-pattern-overview}
C.~Preschern, N.~Kajtazovic, A.~H{\"o}ller, and C.~Kreiner, ``Pattern-based safety development methods: overview and comparison,'' in \emph{Proceedings of the 19th European Conference on Pattern Languages of Programs}, 2014, pp. 1--20.

\bibitem[S73]{S82-principled-assurance}
N.~Annable, M.~Lawford, R.~F. Paige, and A.~Wassyng, ``Principled safety assurance arguments,'' in \emph{International Conference on Computer Safety, Reliability, and Security}.\hskip 1em plus 0.5em minus 0.4em\relax Springer, 2025, pp. 18--32.

\bibitem[S74]{S83-prompting-gpt4}
M.~Sivakumar, A.~B. Belle, J.~Shan, and K.~K. Shahandashti, ``Prompting gpt--4 to support automatic safety case generation,'' \emph{Expert Systems with Applications}, vol. 255, p. 124653, 2024.

\bibitem[S75]{S84-querying-mmint}
A.~Di~Sandro, S.~Kokaly, R.~Salay, and M.~Chechik, ``Querying automotive system models and safety artifacts with mmint and viatra,'' in \emph{2019 ACM/IEEE 22nd international conference on model driven engineering languages and systems companion (MODELS-c)}.\hskip 1em plus 0.5em minus 0.4em\relax IEEE Computer Society, 2019, pp. 2--11.

\bibitem[S76]{S85-ml-assurance-reliability}
Y.~Dong, W.~Huang, V.~Bharti, V.~Cox, A.~Banks, S.~Wang, X.~Zhao, S.~Schewe, and X.~Huang, ``Reliability assessment and safety arguments for machine learning components in system assurance,'' \emph{ACM transactions on embedded computing systems}, vol.~22, no.~3, pp. 1--48, 2023.

\bibitem[S77]{S86-dynamic-open-systems-}
L.~Buysse, I.~Habli, D.~Vanoost, and D.~Pissoort, ``Safe autonomous systems in a changing world: Operationalising dynamic safety cases,'' \emph{Safety Science}, vol. 191, p. 106965, 2025.

\bibitem[S78]{S87-automated-driving-context}
S.~Ballingall, M.~Sarvi, and P.~Sweatman, ``Safety assurance concepts for automated driving systems,'' \emph{SAE International Journal of Advances and Current Practices in Mobility}, vol.~2, no. 2020-01-0727, pp. 1528--1537, 2020.

\bibitem[S79]{S88-offoperational-ml}
H.~Fujino, N.~Kobayashi, and S.~Shirasaka, ``Safety assurance case description method for systems incorporating off-operational machine learning and safety device,'' in \emph{INCOSE International Symposium}, vol.~29.\hskip 1em plus 0.5em minus 0.4em\relax Wiley Online Library, 2019, pp. 152--164.

\bibitem[S80]{S90-autonomous-ml}
C.~Paterson, R.~Hawkins, C.~Picardi, Y.~Jia, R.~Calinescu, and I.~Habli, ``Safety assurance of machine learning for autonomous systems,'' \emph{Reliability Engineering \& System Safety}, vol. 264, p. 111311, 2025.

\bibitem[S81]{S91-chassis-control}
S.~Burton, I.~Kurzidem, A.~Schwaiger, P.~Schleiss, M.~Unterreiner, T.~Graeber, and P.~Becker, ``Safety assurance of machine learning for chassis control functions,'' in \emph{International Conference on Computer Safety, Reliability, and Security}.\hskip 1em plus 0.5em minus 0.4em\relax Springer, 2021, pp. 149--162.

\bibitem[S82]{S92-perception-functions}
S.~Burton, C.~Hellert, F.~H{\"u}ger, M.~Mock, and A.~Rohatschek, ``Safety assurance of machine learning for perception functions,'' in \emph{Deep Neural Networks and Data for Automated Driving: Robustness, Uncertainty Quantification, and Insights Towards Safety}.\hskip 1em plus 0.5em minus 0.4em\relax Springer International Publishing Cham, 2022, pp. 335--358.

\bibitem[S83]{S93-pattern-reuse}
T.~P. Kelly and J.~A. McDermid, ``Safety case construction and reuse using patterns,'' in \emph{Safe Comp 97: The 16th International Conference on Computer Safety, Reliability and Security}.\hskip 1em plus 0.5em minus 0.4em\relax Springer, 1997, pp. 55--69.

\bibitem[S84]{S94-maintenance-review}
C.~C{\^a}rlan, B.~Gallina, and L.~Soima, ``Safety case maintenance: a systematic literature review,'' in \emph{International Conference on Computer Safety, Reliability, and Security}.\hskip 1em plus 0.5em minus 0.4em\relax Springer, 2021, pp. 115--129.

\bibitem[S85]{S95-inability-template}
A.~UK, ``Safety case template for inability arguments,'' AISI UK, Tech. Rep., 2024. [Online]. Available: \url{https://www.aisi.gov.uk/blog/safety-case-template-for-inability-arguments}


\bibitem[S86]{S96-frontier-cyber-inability-}
A.~Goemans, M.~D. Buhl, J.~Schuett, T.~Korbak, J.~Wang, B.~Hilton, and G.~Irving, ``Safety case template for frontier ai: A cyber inability argument,'' \emph{arXiv preprint arXiv:2411.08088}, 2024.

\bibitem[S87]{S97-autonomous-template}
R.~Bloomfield, G.~Fletcher, H.~Khlaaf, L.~Hinde, and P.~Ryan, ``Safety case templates for autonomous systems,'' \emph{arXiv preprint arXiv:2102.02625}, 2021.

\bibitem[S88]{S98-advanced-control}
R.~Alexander, M.~Hall-May, T.~Kelly, and J.~A. McDermid, ``Safety cases for advanced control software: Final report,'' University of York, York, United Kingdom, Final Report, Jun. 2007, approved for public release; distribution unlimited.

\bibitem[S89]{S99-frontier-ai-}
M.~D. Buhl, G.~Sett, L.~Koessler, J.~Schuett, and M.~Anderljung, ``Safety cases for frontier ai,'' \emph{arXiv preprint arXiv:2410.21572}, 2024.

\bibitem[S90]{S100-scalable-frontier}
B.~Hilton, M.~D. Buhl, T.~Korbak, and G.~Irving, ``Safety cases: A scalable approach to frontier {AI} safety,'' \emph{arXiv preprint arXiv:2503.04744}, 2025.

\bibitem[S91]{S101-advanced-ai}
J.~Clymer, N.~Gabrieli, D.~Krueger, and T.~Larsen, ``Safety cases: How to justify the safety of advanced {AI} systems,'' \emph{preprint arXiv:2403.10462}, 2024.

\bibitem[S92]{S102-ai-sil}
S.~Diemert, L.~Millet, J.~Groves, and J.~Joyce, ``Safety integrity levels for artificial intelligence,'' in \emph{International Conference on Computer Safety, Reliability, and Security}.\hskip 1em plus 0.5em minus 0.4em\relax Springer, 2023, pp. 397--409.

\bibitem[S93]{S103-ml-sepsis-design}
Y.~Jia, T.~Lawton, J.~Burden, J.~McDermid, and I.~Habli, ``Safety-driven design of machine learning for sepsis treatment,'' \emph{Journal of Biomedical Informatics}, vol. 117, p. 103762, 2021.

\bibitem[S94]{S104-security-adaptation}
S.~Jahan, A.~Marshall, and R.~Gamble, ``Self-adaptation strategies to maintain security assurance cases,'' in \emph{2018 IEEE 12th International Conference on Self-Adaptive and Self-Organizing Systems (SASO)}.\hskip 1em plus 0.5em minus 0.4em\relax IEEE, 2018, pp. 180--185.

\bibitem[S95]{S106-uncertainty-assurance}
Y.~Matsuno, F.~Ishikawa, and S.~Tokumoto, ``Tackling uncertainty in safety assurance for machine learning: continuous argument engineering with attributed tests,'' in \emph{International Conference on Computer Safety, Reliability, and Security}.\hskip 1em plus 0.5em minus 0.4em\relax Springer, 2019, pp. 398--404.

\bibitem[S96]{S108-big-argument2}
I.~Habli, R.~Hawkins, C.~Paterson, P.~Ryan, Y.~Jia, M.~Sujan, and J.~McDermid, ``The big argument for {AI} safety cases,'' \emph{arXiv preprint arXiv:2503.11705}, 2025.

\bibitem[S97]{S109-perception-link}
R.~Salay, K.~Czarnecki, H.~Kuwajima, H.~Yasuoka, V.~Abdelzad, C.~Huang, M.~Kahn, V.~D. Nguyen, and T.~Nakae, ``The missing link: Developing a safety case for perception components in automated driving,'' \emph{SAE International Journal of Advances and Current Practices in Mobility}, vol.~5, no. 2022-01-0818, pp. 567--579, 2022.

\bibitem[S98]{S110-systematic-machinery}
F.~Wolny, S.~Vock, T.~Holoyad, and R.~Adler, ``The need for systematic approaches in risk assessment of safety-critical ai-applications in machinery,'' in \emph{European Safety and Reliability \& Society for Risk Analysis Europe Conference ESREL SRA-E}, 2025.

\bibitem[S99]{S111-autonomy-framework}
M.~Wagner and C.~Carlan, ``The open autonomy safety case framework,'' \emph{arXiv preprint arXiv:2404.05444}, 2024.

\bibitem[S100]{S112-certification-regulation}
N.~Leveson, ``The use of safety cases in certification and regulation,'' \emph{Journal of System Safety}, 2011, aeronautics and Astronautics/Engineering Systems, MIT.

\bibitem[S101]{S113-asl4-sketches}
R.~Grosse, ``Three sketches of asl-4 safety case components,'' \emph{Anthropic Alignment Science Blog}, 2024.

\bibitem[S102]{S114-harmonized-argument}
M.~Loba, N.~F. Salem, M.~Nolte, A.~Dotzler, D.~Ludwig, and M.~Maurer, ``Toward a harmonized approach--requirement-based structuring of a safety assurance argumentation for automated vehicles,'' \emph{arXiv preprint arXiv:2505.03709}, 2025.

\bibitem[S103]{S115-learning-cps}
M.~Bagheri, J.~Lamp, X.~Zhou, L.~Feng, and H.~Alemzadeh, ``Towards developing safety assurance cases for learning-enabled medical cyber-physical systems,'' \emph{arXiv preprint arXiv:2211.15413}, 2022.

\bibitem[S104]{S116-evaluation-ai-scheming}
M.~Balesni, M.~Hobbhahn, D.~Lindner, A.~Meinke, T.~Korbak, J.~Clymer, B.~Shlegeris, J.~Scheurer, C.~Stix, R.~Shah \emph{et~al.}, ``Towards evaluations-based safety cases for {AI} scheming,'' \emph{arXiv preprint arXiv:2411.03336}, 2024.

\bibitem[S105]{S117-goal-based-certification}
E.~Stensrud, T.~Skramstad, J.~Li, and J.~Xie, ``Towards goal-based software safety certification based on prescriptive standards,'' in \emph{2011 First International Workshop on Software Certification}.\hskip 1em plus 0.5em minus 0.4em\relax IEEE, 2011, pp. 13--18.

\bibitem[S106]{S118-trusta-formal-llm}
Z.~Chen, Y.~Deng, and W.~Du, ``Trusta: Reasoning about assurance cases with formal methods and large language models,'' \emph{Science of Computer Programming}, vol. 244, p. 103288, 2025.

\bibitem[S107]{S119-ul4600-autonomous}
P.~Koopman, ``Ul 4600: What to include in an autonomous vehicle safety case,'' \emph{Computer}, vol.~56, no.~05, pp. 101--104, 2023.

\bibitem[S108]{S120-uncertainty-pattern}
Y.~Idmessaoud, J.-L. Farges, E.~Jenn, V.~Mussot, A.~F. Pires, F.~Chenevier, and R.~C. Laguna, ``Uncertainty in assurance case pattern for machine learning,'' in \emph{Embedded Real Time Systems (ERTS)}, 2024.

\bibitem[S109]{S121-upstream-downstream}
J.~McDermid, Y.~Jia, and I.~Habli, ``Upstream and downstream {AI} safety: Both on the same river?'' \emph{arXiv preprint arXiv:2501.05455}, 2024.

\bibitem[S110]{S122-gpt4-turbo-defeaters}
K.~K. Shahandashti, A.~B. Belle, M.~M. Mohajer, O.~Odu, T.~C. Lethbridge, H.~Hemmati, and S.~Wang, ``Using gpt-4 turbo to automatically identify defeaters in assurance cases,'' in \emph{2024 IEEE 32nd International Requirements Engineering Conference Workshops (REW)}.\hskip 1em plus 0.5em minus 0.4em\relax IEEE, 2024, pp. 46--56.

\bibitem[S111]{S123-autonomous-trust-public}
T.~Myklebust, T.~St{\aa}lhane, G.~D. Jenssen, and I.~W{\ae}r{\o}, ``Autonomous cars, trust and safety case for the public,'' in \emph{2020 Annual Reliability and Maintainability Symposium (RAMS)}.\hskip 1em plus 0.5em minus 0.4em\relax IEEE, 2020, pp. 1--6.

\bibitem[S112]{S124-safe-reinforcement}
J.~Bragg and I.~Habli, ``What is acceptably safe for reinforcement learning?'' in \emph{International Conference on Computer Safety, Reliability, and Security}.\hskip 1em plus 0.5em minus 0.4em\relax Springer, 2018, pp. 418--430.
}
\end{thebibliography}

\end{document}